%% file: main.tex
\documentclass[11pt,a4paper]{article}
\pdfoutput=1

\input{packages.tex}
\input{macros.tex}

\usepackage{empheq}                                              
\newcommand*\widefbox[1]{\fbox{\hspace{1em}#1\hspace{1em}}}      
\usepackage{cleveref}
\usepackage[super]{nth}
\usepackage{xspace}
\usepackage[math]{cellspace}

\newcommand{\F}{\mathcal{F}}
\newcommand{\G}{\mathcal{G}}
\newcommand{\FF}{\mathbb{F}_{\rm LOS}}
\newcommand{\GG}{\mathbb{G}_{\rm LOS}}
\newcommand{\T}{\underline\theta}

\newcommand{\lenstronomy}{\texttt{lenstronomy}\xspace}

\title{Weak lensing of strong lensing: beyond the tidal regime}

\author[a]{Théo Duboscq,}
\author[a]{Natalie B. Hogg,}
\author[a, b]{Pierre Fleury,}
\author[a]{Julien Larena}

\affiliation[a]{Laboratoire Univers et Particules de Montpellier (LUPM), 
CNRS \& Université de Montpellier (UMR-5299),
Parvis Alexander Grothendieck, F-34095 Montpellier Cedex 05, France}

\affiliation[b]{Universit\'{e} Paris-Saclay, CNRS, CEA, Institut de physique th\'{e}orique, 91191, Gif-sur-Yvette, France}

\emailAdd{theo.duboscq@umontpellier.fr}
\emailAdd{natalie.hogg@lupm.in2p3.fr}
\emailAdd{pierre.fleury@lupm.in2p3.fr}
\emailAdd{julien.larena@umontpellier.fr}

\abstract{The analysis of strong lensing images usually involves an external convergence and shear, which are meant to model the effect of perturbations along the line of sight, on top of the main lens. Such a description of line-of-sight perturbations supposes that the corresponding gravitational fields can be treated in the tidal regime. Going one step further introduces additional effects, known as flexion, which have been hitherto neglected in strong lensing. In this work, we build a minimal model for the line-of-sight flexion, which adds four new complex parameters to the lens model. Contrary to convergence and shear, the line-of-sight flexion cannot be projected onto the main lens plane. For a $\Lambda$CDM cosmology, we predict the typical line-of-sight flexion to be on the order of $10^{-3}\,\mathrm{arcsec}^{-1}$ on galactic scales. Neglecting its effect in lens modelling is found to bias the recovery of other parameters; in particular, the line-of-sight shear can be biased up to $2\sigma$. Accounting for the line-of-sight flexion in our minimal framework restores accuracy, at the the cost of degrading precision. With current imaging capabilities, the line-of-sight flexion is unlikely to be measurable on individual strong lensing images; it must therefore be considered a nuisance parameter rather than an observable in its own right.
}


\date{\today}

\begin{document}

\maketitle
\flushbottom

\section{Introduction}
\label{Intro}

Gravitational lensing -- the deflection of light by massive objects -- is a central tool to explore the properties of our Universe. Correlations in the positions and shapes of galaxies induced by the weak gravitational lensing of their light by foreground galaxies allow us to map the large scale distribution of matter in the Universe. Weak lensing measurements have been used to constrain key ingredients of the cosmological model, such as the matter density $\Omega_{\rm m}$ and the amount of clustering in the Universe, quantified by $\sigma_8$, the amplitude of the linear matter power spectrum on a scale of $8$ Mpc/$h$. 

The Kilo-Degree Survey (KiDS) and the Dark Energy Survey (DES), which in tandem mapped nearly $5000$ deg$^2$ of the sky, provided a joint constraint on $S_8 = \sigma_8 \sqrt{\Omega_{\rm m}/0.3} = 0.790^{+0.018}_{-0.014}$ \cite{Kilo-DegreeSurvey:2023gfr}, in good agreement with the value derived from the cosmic microwave background power spectra as measured by the \textit{Planck} satellite \cite{Planck2018}. Furthermore, weak lensing measurements are one of the key science goals of a number of ongoing or forthcoming surveys, such as \textit{Euclid}, the Roman Space Telescope and the Vera C. Rubin Observatory.


On the scales of individual galaxies and galaxy clusters, the phenomenon of strong lensing may be observed. When the source is variable, such as a quasar or supernova, a geometric measurement of the Hubble parameter at redshift zero, $H_0$, can be made. This technique was first proposed by Refsdal \cite{Refsdal1964}, and subsequently performed by the H0LiCOW and TDCOSMO collaborations \cite{Wong:2019kwg, Millon:2019slk}. For example, the first TDCOSMO constraint on $H_0$ using six strongly lensed quasars found $H_0 = 74.0^{+1.7}_{-1.8}$ km s$^{-1}$ Mpc$^{-1}$ \cite{Millon:2019slk}. An updated analysis, including an additional strongly lensed quasar and stellar kinematics data to avoid reliance on a strict mass profile for the lens galaxies, found $H_0 = 67.4^{+4.1}_{-3.2}$ km s$^{-1}$ Mpc$^{-1}$ \cite{Birrer:2020tax}, which is in agreement with the measurement derived from the cosmic microwave background within the framework of the standard cosmological model, $\Lambda$CDM. 



Furthermore, careful modelling of the lensing event may allow for the identification of individual dark subhaloes inside the main lens or on the line of sight connecting source and observer, manifesting as weak perturbers of the strong lensing image generated by the compact, strong lens. These haloes, too small to contain any significant amount of visible matter, constitute a direct probe of the properties of dark matter on kiloparsec length scales and $10^{7}$ to $10^{9} M_{\odot}$ mass scales \cite{Vegetti2010}. Sufficient detections of individual subhaloes will allow their abundance to be compared to a theoretical subhalo mass function, which in turn may bolster or weaken the evidence for the weakly interacting massive particle model of dark matter \cite{Cyr-Racine:2015jwa, Nightingale:2022bhh}; see \cite{Vegetti:2023mgp} for a recent review.

However, the impact of individual subhaloes on images is very small and only detectable if the subhalo is located in the vicinity of the event's critical curve. Alternatively, we may be able to characterise the collective effects of a population of such subhaloes on images, hoping for an enhanced  signal \cite{He:2021rjd, CaganSengul:2020nat}. Notably, unlike individual subhaloes, it seems that these population effects can be separated into contributions from line-of-sight (LOS) haloes located outside the strong lens and small perturbing haloes within the lens galaxy itself.

The effect of LOS perturbations on strong lensing images was proposed by Birrer et al. to be a potential cosmological probe in its own right \cite{Birrer:2016xku, Birrer:2017sge}. At first order, LOS perturbers induce shear on a strong lensing image, analogous to the cosmic shear measured in weak lensing surveys. If measureable, this shear could provide an independent constraint on, for example, $S_8$. Fleury et al. \cite{LOS_in_SGL} produced a comprehensive treatment of LOS effects in strong lensing, deriving a minimal model for the LOS shear, and demonstrating that the parameters of this model escape degeneracies with those of the main lens, which is crucial for the use of LOS shear for cosmology. The success of the minimal model was demonstrated by Hogg et al. \cite{Hogg_2023}, who also showed that the LOS shear is systematically measureable from mock strong lensing images with high precision.

However, these results were obtained under the assumption of the tidal regime, meaning that perturbations to the strong lens can be modelled as tidal fields. If this were not the case, the beyond-shear effect of flexion \cite{Goldberg:2004hh, Bacon_2006, Clarkson_2015} -- which produces skewing and triangularisation -- could be present in a strong lensing image. Using a proxy for the flexion, Hogg et al. investigated this assumption and found that a significant number of dark matter haloes in a simulated line of sight may indeed violate the tidal approximation, but that LOS shear measurements remain feasible.

The aim of the current work is twofold: firstly, we extend the LOS formalism to the flexion regime, deriving a minimal flexion model. This has a similar motivation to the derivation of the minimal shear model previously obtained by Fleury et al.; we wish to model the effect of LOS flexion with as few parameters as possible, and with parameters that will not be degenerate with other quantities of interest. Secondly, we aim to quantify the expected cosmic flexion signal, and whether the presence of flexion in a strong lensing image can affect the accuracy and precision of an attempted shear measurement. Our newly derived minimal flexion model will enable us to perform these tests in a much more robust way than the simplistic proxy for flexion used in \cite{Hogg_2023}. This is a key step, since the mis-modelling of flexion has already been shown to bias $H_0$ measurements high \cite{Teodori:2023nrz}.


This paper is organised as follows: in \cref{sec:LOStheory}, we review the LOS formalism for strong lensing as first developed in \cite{LOS_in_SGL}. In \cref{Section : Minimal flexion model} we derive the so-called minimal model for flexion, a model which encodes the effects of LOS flexion on a strong lensing image with the minimum number of parameters. In \cref{Section: LOS flexion} we address the measurability of flexion, computing the expected variance of cosmic flexion in $\Lambda$CDM, and demonstrate the advantage of the minimal model in fitting a mock strong lensing image. In \cref{Sec: LOS shear measure}, we stress-test the measurability of LOS shear in the presence of LOS flexion, finding that when flexion is neglected, shear measurements can indeed be jeopardised. Lastly, we present our conclusions in \cref{sec:conclusion}. Throughout the article, we assume that the background spacetime through which light propagates is described by the Friedmann--Lemaître--Robertson--Walker (FLRW) model.

\section{LOS effects up to flexion} \label{sec:LOStheory}

In this section we review the theoretical background needed to introduce the minimal flexion model in \cref{Section : Minimal flexion model}.

\subsection{Partial displacement, convergence, shear and flexion}\label{Def of quantities}

In the simplest strong lensing scenario, light rays are deflected by a single thin lens. This means that we define a lens plane upon which we project the mass of the lens, perpendicular to a given optical axis. In this situation, the problem is described by the lens equation,
\begin{equation}
    \boldsymbol\beta = \boldsymbol\theta - \boldsymbol\alpha(\boldsymbol\theta) \;.
\end{equation}
In this equation, $\boldsymbol\beta$ is the angular position of the source with respect to the optical axis, $\boldsymbol\theta$ is the angular position of the light ray when it crosses the lens plane, and $\boldsymbol\alpha(\boldsymbol\theta)$ is the lens displacement angle which is related to the gravitational potential associated with the surface mass density of the lens across the lens plane.

In a more physically realistic situation, the observer, lens and source will not be the only objects in the Universe; many thousands of galaxies may lie along a given line of sight. One can construct a more elaborate model to include the effects of such objects. We could consider $N$ masses along the line of sight, each being modelled by a thin lens. This situation can be described using the multi-plane formalism \citep{Blandford_Narayan1986}. However, in this case the lens equation becomes a recursive relation which makes analytical computation intractable. 

\begin{figure}[h]
    \centering
    \includegraphics[width=\textwidth]{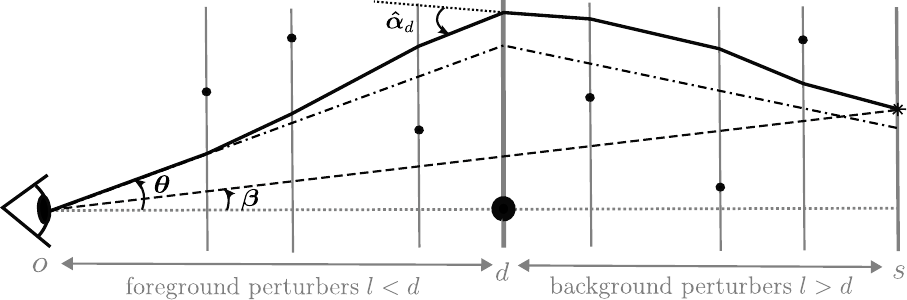}
    \caption{Multi-plane lensing in the dominant lens approximation. The total displacement is dominated by the effect of the main lens (d) whilst the other lenses are treated as perturbers. The optical axis (dotted line) is conventionally aligned with the centre of mass of the dominant lens. The thick solid line represents the physical ray; the dashed line shows the path of the ray if no lenses were present; the dot-dashed line indicates the dominant-lens-only ray.}
    \label{Fig: Dominant lens setup}
\end{figure}

\begin{figure}
    \centering
    \includegraphics[width=\textwidth]{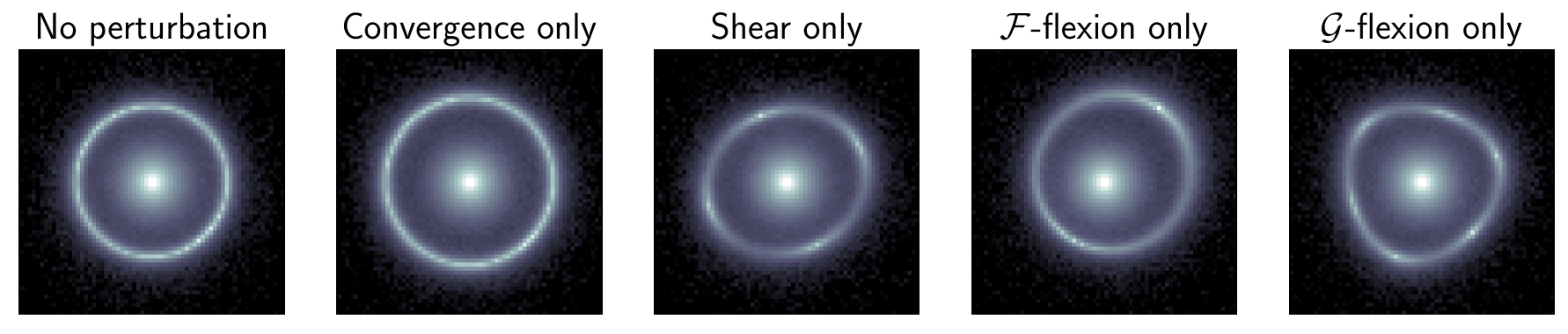}
    \caption{Effect of the different lensing distortions on an Einstein ring produced by a singular isothermal sphere. For the purpose of illustration, the values of LOS convergence, shear, type $\F$-flexion and type $\G$-flexion are exaggerated compared to what we expect for a realistic lens.}
    \label{fig: illustration deform}
\end{figure}

A possible solution to this problem is to consider a dominant lens as the main contributor to the deflection of light rays, whilst inhomogeneities along the line of sight are treated as perturbers. This is the dominant lens formalism developed in \cite{LOS_in_SGL}. The situation is depicted in \cref{Fig: Dominant lens setup}, where the dominant lens is labelled by the index $d$, the observer plane (plane 0) with $o$ and the source plane (plane $N+1$) with $s$. 
Then, for $(i,j)\in [\![0,N+1]\!]^2$, we call $D_{ij}$ the angular diameter distance to plane $j$ as seen from plane $i$. In the FLRW cosmology, $D_{ij} = f_K(\chi_j - \chi_i)/(1+z_j)$, where $z_j$ is the cosmological redshift of $j$, $\chi_i$ and $\chi_j$ are the comoving distances of $i$ and $j$ from the observer, and $f_K$ is defined by
\begin{equation}
f_K(\chi) 
\equiv
\begin{cases}
\sin\left(\sqrt{K}\chi\right)/\sqrt{K} & \text{if }K > 0, \\
\chi & \text{if }K = 0, \\
\sinh\left(\sqrt{-K}\chi\right)/\sqrt{-K} & \text{if }K > 0,
\end{cases}
\end{equation}
with $K$ the FLRW spatial curvature parameter. Each lens plane $l$ is characterised by its surface mass density, from which we define a potential $\hat{\psi}_l$. The quantity
\begin{equation}
    \hat{\underline\alpha}_l (\underline{x_l}) \equiv 2 \frac{\partial\hat{\psi}_l}{\partial \underline x_l^*}(\underline x_l)
\end{equation}
is called the deflection angle, where $\underline x_l$ is the transverse position in the $l$-th 
plane. The deflection angle is defined in the rest frame of the lens. Here, we used complex notation, in which we replace two-dimensional vectors by their complex counterpart.\footnote{Note that within this formalism, all the functions we will consider must be treated as functions of two independent variables, which are a complex number and its conjugate, meaning that we should write $\hat{\underline\alpha}_l (\underline{x_l}, \underline x_l^*)$. However, we will omit the conjugate in the relevant arguments to keep the expressions shorter. For more details about the complex formalism, we refer the reader to subsubsec. 2.3.2 of \cite{LOS_in_SGL}.} In this context, we define the complex derivatives as
\begin{align}
    \frac{\partial}{\partial \underline x_l} \equiv \frac{1}{2} \left( \frac{\partial}{\partial x_{l,1}} - \mathrm{i}\frac{\partial}{\partial x_{l,2}} \right) \;, \hspace{2em} \frac{\partial}{\partial \underline x_l^*} \equiv \frac{1}{2} \left( \frac{\partial}{\partial x_{l,1}} + \mathrm{i}\frac{\partial}{\partial x_{l,2}} \right) \;.
\end{align}
For $i<j$, we call
\begin{equation}
    \underline\beta_{ij} \equiv \frac{\underline x_j}{D_{ij}}
\end{equation}
the angular direction in which we would see $\underline x_j$ from the $i$-th plane without intermediate lenses. From there, we can define partial displacement, convergence, shear and flexion. Consider now three indices $i< l< j$. Then the partial displacement angle is defined by\begin{equation}
    \underline \alpha_{ilj}(\underline\beta_{il}) \equiv \frac{D_{lj}}{D_{ij}} \hat{\underline\alpha}_l (\underline x_l) \;.
\end{equation}
This displacement angle corresponds to the displacement an observer in the $i$-th plane would see with only one lens located in the $l$-th plane and a source in $j$-th plane. For instance, $\underline\alpha_{ods}$ corresponds to the deflection angle of the main lens alone. The partial convergence and shear are then defined as $\kappa_{ilj} \equiv \partial \underline\alpha_{ilj}/\partial \underline\beta_{il}$ and $\gamma_{ilj} \equiv \partial \underline\alpha_{ilj}/\partial \underline\beta^*_{il}$. Convergence and shear are thus the second order derivatives of the potential, up to some distance factor, namely \begin{equation}
    \kappa_{ilj} = 2\frac{D_{il}D_{lj}}{D_{ij}} \frac{\partial^2 \hat{\psi}_l}{\partial\underline x_l\partial\underline x_l^*}  \;, \qquad \gamma_{ilj} = 2\frac{D_{il}D_{lj}}{D_{ij}} \frac{\partial^2 \hat{\psi}_l}{\partial\underline x_l^{*2}} \;.
\end{equation}
The partial convergence and shear are once again the convergence and shear an observer in the $i$-th plane would measure with only one lens in the $l$-th plane and a source in the $j$-th plane. Finally, the flexion parameters are linked to the third derivatives of the potential. We define
\begin{equation}
    \hat{\mathcal{F}}_l \equiv 2\frac{\partial}{\partial\underline x_l} \frac{\partial}{\partial\underline x_l^*} \frac{\partial}{\partial\underline x_l^*} \hat{\psi}_l \;, \qquad
    \hat{\mathcal{G}}_l \equiv 2\frac{\partial}{\partial\underline x_l^*} \frac{\partial}{\partial\underline x_l^*} \frac{\partial}{\partial\underline x_l^*} \hat{\psi}_l \;,
\end{equation}
where $\hat{\mathcal{F}}_l$ and $\hat{\mathcal{G}}_l$ are the two flexion types. Here there are only two independent complex flexion parameters, because the other third derivatives of the potential are just the conjugates of the above definitions. The effect of these quantities on an Einstein ring produced by a simple lens model consisting of a Singular Isothermal Sphere (SIS, \cite{Kormann1994}) is shown in \cref{fig: illustration deform}.
To generate those images, we used the software \lenstronomy \cite{Birrer:2018xgm, Birrer2021}, in conjunction with the LOS flexion formalism developed in section \cref{Section : Minimal flexion model}, where we define the notion of LOS convergence, shear and flexion. For each panel, we put a non zero magnitude for one of these LOS parameters. An implementation of this model in \lenstronomy is available on Github in a branch called \texttt{los-flexion}.\footnote{\url{https://github.com/TheoDuboscq/lenstronomy/tree/los-flexion}.}
We then define the partial flexion parameters, which can be linked to the ones defined above, as 
\begin{align}
    \mathcal{F}_{ilj} &\equiv \frac{\partial \gamma_{ilj}}{\partial\underline\beta_{il}} = \frac{\partial \kappa_{ilj}}{\partial\underline\beta_{il}^*} = \frac{D_{il}^2D_{lj}}{D_{ij}}\hat{\mathcal{F}}_l \;, \\
    \mathcal{G}_{ilj} &\equiv \frac{\partial \gamma_{ilj}}{\partial\underline\beta_{il}^*} = \frac{D_{il}^2D_{lj}}{D_{ij}}\hat{\mathcal{G}}_l \;.
\end{align}
Again, the partial flexion may be understood as the flexion an observer in the $i$-th plane would measure with only one lens in the $l$-th plane and a source in the $j$-th plane. 

The dominant lens approximation consists in assuming that all lensing quantities associated with the secondary lenses are small compared to those coming from the dominant lens and can therefore be treated perturbatively. Namely, if we introduce a small power-counting parameter $\epsilon$, every lensing quantity $q$ satisfies the following: $\forall \; l\neq d, \; \forall \; i<l<j, \; |q_{ilj}| = O(\epsilon)$. We also assume that linear combinations of these small lensing quantities are still perturbations. These assumptions allow one to derive a lens equation which does not involve a recursion relation and which is exclusively expressed in terms of $\underline\theta \equiv \underline\beta_{od}$ (see \cite{LOS_in_SGL} for details):
\begin{align}\label{Dominant lens LE}
    \underline\beta &= \underline\theta - \underline\alpha(\underline\theta) \;, \\
    \underline\alpha(\underline\theta) &= \underline\alpha_{ods}\big[\underline\theta - \underline\alpha_{od}(\underline\theta) \big] + \sum_{l<d} \underline\alpha_{ols}(\underline\theta) + \sum_{l>d} \underline\alpha_{ols} \big[\underline\theta - \underline\alpha_{odl}(\underline\theta)\big] + O(\epsilon^2),
\end{align}
with $\underline\alpha_{od}(\underline\theta) = \sum_{l<d} \underline\alpha_{old}(\underline\theta)$ being the displacement of a fictitious source in the plane of the deflector due to the foreground lenses only.

\subsection{The flexion regime}

In \cite{LOS_in_SGL}, the authors then apply the tidal approximation, which consists in assuming that the shear and convergence fields of the perturbers are homogeneous across the relevant part of the sky. Here, we will go one step further, allowing for gradients of shear and convergence, but assuming that flexion is homogeneous across the image. Using the definitions presented in \cref{Def of quantities}, one finds
\begin{equation}
    \underline\alpha_{ilj}(\underline\theta) = \underline\alpha_{ilj}(0) + \frac{D_{ol}}{D_{il}}\kappa_{ilj}(0)\underline\theta + \frac{D_{ol}}{D_{il}}\gamma_{ilj}(0)\underline\theta^* +\frac{D_{ol}^2}{2D_{il}^2} \left( \mathcal{F}_{ilj}^*\underline\theta^2 + 2\mathcal{F}_{ilj}|\underline\theta|^2 + \mathcal{G}_{ilj}\underline\theta^{*2}\right) \;.
\end{equation}
Plugging this into the general lens equation \cref{Dominant lens LE}, and after reorganising all the terms, this yields
\begin{equation}\label{alpha}
\begin{split}
    \underline\alpha(\underline\theta) =& \; \underline\alpha_{ods}\left\{(1-\kappa_{od})\underline\theta - \gamma_{od}\underline\theta^* - \frac{1}{2}\left[ \mathcal{F}_{od}^*\underline\theta^2 + 2\mathcal{F}_{od}\underline\theta \underline\theta^* + \mathcal{G}_{od} (\underline\theta^*)^2 \right] - \underline\alpha_{od}(0) \right\}  \\
    &+ \underline\alpha_{os}(0) + \kappa_{os}\underline\theta + \gamma_{os}\underline\theta^* + \frac{1}{2} \left[\mathcal{F}_{os}^*\underline\theta^2 + 2\mathcal{F}_{os}\underline\theta\underline\theta^* + \mathcal{G}_{os}(\underline\theta^*)^2\right] \\
    &- \left[\kappa_{ds}+\prescript{(2)}{}{\mathcal{F}}_{ds}^*\underline\theta +  \prescript{(2)}{}{\mathcal{F}}_{ds}\underline\theta^* \right] \underline\alpha_{ods} (\underline\theta) - \left[ \gamma_{ds}+\prescript{(2)}{}{\mathcal{F}}_{ds}\underline\theta +  \prescript{(2)}{}{\mathcal{G}}_{ds}\underline\theta^* \right] \underline\alpha^*_{ods} (\underline\theta)  \\
    &+ \frac{1}{2} \left\{ \prescript{(1)}{}{\F}^*_{ds}\underline\alpha^2_{ods}(\underline\theta) + 2\prescript{(1)}{}{\F}_{ds}\underline\alpha^*_{ods}(\underline\theta) \underline\alpha_{ods}(\underline\theta)  + \prescript{(1)}{}{\G}_{ds}[\underline\alpha^*_{ods}(\underline\theta)]^2 \right\},
\end{split}
\end{equation}

\noindent
at first order in $\epsilon$. Here, the definitions of the various quantities involved are the following:
\begin{align}
    &\kappa_{od} \equiv \sum_{l<d} \kappa_{old}(0) \;, \hspace{1em} \kappa_{os} \equiv \sum_{l\neq d} \kappa_{ols}(0) \;, \hspace{1em} \kappa_{ds} \equiv \sum_{l>d} \kappa_{dls}(0) \;, \\
    \mathcal{F}_{od} \equiv \sum_{l<d} & \mathcal{F}_{old}\;, \hspace{1em} \mathcal{F}_{os} \equiv \sum_{l\neq d} \mathcal{F}_{ols}\;, \hspace{1em} \prescript{(1)}{}{\F}_{ds} \equiv \sum_{l>d} \frac{D_{os}}{D_{ds}} \mathcal{F}_{dls} \;, \hspace{1em} \prescript{(2)}{}{\mathcal{F}}_{ds} \equiv \sum_{l>d} \frac{D_{ol}}{D_{dl}} \mathcal{F}_{dls} \;,
\end{align}
whilst the shear quantities are defined in the same way as the convergence quantities, and the type-$\mathcal{G}$ flexion quantities are defined like the type-$\mathcal{F}$ flexion ones. As explained in detail in~\cite{LOS_in_SGL}, the $\underline\alpha_{od}(0)$ and $\underline\alpha_{os}(0)$ terms in \cref{alpha} can be removed, since they are homogeneous displacements that are not measurable in a given image. The $\underline\alpha_{od}(0)$ term can be absorbed by shifting the origin of the lens plane, and $\underline\alpha_{os}(0)$ will disappear thanks to a redefinition of the origin of the source plane. Therefore, we shall simplify the expression of $\underline\alpha$ by omitting those terms.


\section{Dealing with degeneracies: the minimal flexion model} \label{Section : Minimal flexion model}

As in the case of the tidal regime developed in \cite{LOS_in_SGL}, there are degeneracies between parameters involved in \cref{alpha}. Indeed, since the source position and shape are unknown, two different parameterisations of the lens and the source that give rise to the same lensing image cannot be distinguished. This is called the mass-sheet degeneracy \cite{Falco1985, Schneider_Sluse2013}. This means that there is some freedom in the values of the parameters of a lens model: one may be able to change the values of two parameters as long as a source transformation allows one to recover the same image. Thus, one will not be able to accurately measure those parameters with a given image, as they are degenerate with one another. The aim of this section is thus to exploit these degeneracies to build a minimal lens model with minimal degeneracies. What is done here generalises what was done at the level of the shear in \cite{LOS_in_SGL}, and the approach is similar. This also means that we recover the results from \cite{LOS_in_SGL} by taking the flexion parameters to zero. To build the minimal model, we need two transformations, one which will affect the main lens model and the other the source.

\subsection{Main lens transformation}
To make the expression shorter, it is convenient to define a function $f$ as 
\begin{align}
    f \colon \underline \theta \mapsto  (1-\kappa_{od})\underline\theta - \gamma_{od}\underline\theta^* - \frac{1}{2}\left[ \mathcal{F}_{od}^*\underline\theta^2 + 2\mathcal{F}_{od}\underline\theta \underline\theta^* + \mathcal{G}_{od} (\underline\theta^*)^2 \right] \;.
\end{align}
Some parameters are degenerate with the main lens. Just as the effect of the foreground shear is degenerate with the main lens ellipticity (see \cite{LOS_in_SGL}), we expect the effect of $\mathcal{F}_{od}$ and $\mathcal{G}_{od}$ to be degenerate with the main lens model. We therefore have to define an effective potential that can absorb the $od$ terms: 
\begin{equation} \label{psi_eff}
    \psi_{\rm eff}(\underline\theta) = \psi_{ods} \left[f(\underline\theta)\right]  \;.
\end{equation}
This defines a new main lens with a different set of parameters. For instance, suppose that $\psi_{ods}$ describes an elliptical Sérsic lens \cite{Cardone:2003iz} (the ellipticity being introduced in the potential, not the convergence). Then, the effect of the foreground shear $\gamma_{od}$ is degenerate with the ellipticity of the Sérsic, which means that $\psi_{ods}(\underline\theta - \gamma_{od}\underline\theta^*)$ defines a new Sérsic with modified ellipticity. In the same vein, $\kappa_{od}$ changes the effective Einstein radius of the lens. 

For the foreground flexion, the procedure is more complicated. Simple lens models like Sérsics have no intrinsic flexion; therefore $\psi_{ods}[f(\underline\theta)]$ no longer defines an elliptical Sérsic. However, for more complex lens models, the foreground flexion \textit{can} be degenerate with the lens model parameters. For instance, if one considers a main lens composed of a baryonic and a dark matter components, then there is a degeneracy between the effect of the foreground type-$\mathcal{F}$ flexion $\mathcal{F}_{od}$ and the effect of the offset between the centres of the two components. This is similar to the case of shear: $\gamma_{od}$ is degenerate with the ellipticity of a Singular Isothermal Ellipsoid (SIE, \cite{Kormann1994}), but of course such degeneracy would be artificially broken if one assumed an overly simple model, e.g. the SIS. 

In summary, more complexity in the description of the line of sight requires more complexity in the description of the main lens, so as to fully account for degeneracies. When degeneracies are fully accounted for, parameter inference can be made for a given lens model and data without fear of loss of accuracy or precision in the measured parameters.

Bearing the above in mind, along with the effective potential defined in \cref{psi_eff}, and defining $\underline\alpha_{\rm eff} \equiv 2\partial \psi_{\rm eff}/\partial\theta^*$,  we can write 
\begin{equation}
\begin{split}
\label{alpha eff}
    \underline\alpha_{\rm eff}(\underline\theta) =& \; \left(1 - \kappa_{od} - \mathcal{F}^*_{od}\underline\theta - \mathcal{F}_{od}\underline\theta^*\right) \alpha_{ods} \left[ f(\underline\theta) \right] - \left(\gamma_{od} + \mathcal{F}_{od}\underline\theta + \mathcal{G}_{od}\underline\theta^* \right) \underline\alpha^*_{ods}\left[ f(\underline\theta) \right]  \\
    =& \alpha_{ods} \left[ f(\underline\theta) \right] - \left(\kappa_{od} + \mathcal{F}^*_{od}\underline\theta + \mathcal{F}_{od}\underline\theta^*\right) \underline\alpha_{ods}(\underline\theta) - \left( \gamma_{od} + \mathcal{F}_{od}\underline\theta + \mathcal{G}_{od}\underline\theta^* \right) \underline\alpha^*_{ods}(\underline\theta) \;.
\end{split}
\end{equation}

\noindent
We can thus rewrite $\underline\alpha$ in terms of $\underline\alpha_{\rm eff}$. Keeping in mind that LOS parameters are assumed to be perturbations, we get
\begin{equation}
\begin{split}
    \underline\alpha(\underline\theta) =& \; \underline\alpha_{\rm eff}(\underline\theta) + \kappa_{os}\underline\theta + \gamma_{os}\underline\theta^* + \frac{1}{2} \left\{\mathcal{F}_{os}^* \underline\theta^2 + 2 \mathcal{F}_{os} \underline\theta \, \underline\theta^* + \mathcal{G}_{os} (\underline\theta^*)^2\right\} \\
    &-  \left\{ \kappa_{ds} - \kappa_{od} + [\prescript{(2)}{}{\mathcal{F}}_{ds}^* - \mathcal{F}_{od}^*]\underline\theta + [\prescript{(2)}{}{\mathcal{F}}_{ds} - \mathcal{F}_{od}]\underline\theta^* \right\} \underline\alpha_{\rm eff}(\underline\theta) \\ 
    &- \left\{ \gamma_{ds} - \gamma_{od} + [\prescript{(2)}{}{\mathcal{F}}_{ds} - \mathcal{F}_{od}]\underline\theta + [\prescript{(2)}{}{\mathcal{G}}_{ds} - \mathcal{G}_{od}]\underline\theta^* \right\} \underline\alpha^*_{\rm eff}(\underline\theta) \\ 
    &+ \frac{1}{2} \left\{ \prescript{(1)}{}{\F}^*_{ds}\underline\alpha^2_{\rm eff}(\underline\theta) + 2\prescript{(1)}{}{\F}_{ds}\underline\alpha^*_{\rm eff}(\underline\theta) \underline\alpha_{\rm eff}(\underline\theta)  + \prescript{(1)}{}{\G}_{ds}[\underline\alpha^*_{\rm eff}(\underline\theta)]^2 \right\} \;.
\end{split}
\end{equation}
Here we have been able to reduce the number of flexion parameters from eight to six by redefining the potential of the main lens. This means that we cannot distinguish between the situation where the main lens is modelled by $\psi_{ods}$ and obeys a lens equation involving eight flexion parameters, and the situation where the main lens is described by $\psi_{\rm eff}$ and obeys a lens equation with a different line of sight, where only six flexion parameters are non-vanishing. 

\subsection{Source-position transformation} Furthermore, since the source position and intrinsic shape are unknown, one can perform what is called a source-position transformation (SPT, \cite{Schneider_2014, Wagner_2018}). We can then use the SPT to spot the degeneracies between the LOS parameters and the source model. We proceed to the following source-position transformation:
\begin{equation} \label{SPT}
\begin{split}
    \Tilde{\underline\beta} \equiv& \; (1- \kappa_{od} + \kappa_{ds})\underline\beta + (\gamma_{ds} - \gamma_{od})\underline\beta^*  \\
    &+ \frac{1}{2}\left[ (\prescript{(2)}{}{\mathcal{F}}_{ds}^* - \mathcal{F}^*_{od}) \underline\beta^2 + 2 (\prescript{(2)}{}{\mathcal{F}}_{ds} - \mathcal{F}_{od}) \underline\beta \underline\beta^* + (\prescript{(2)}{}{\mathcal{G}}_{ds} - \mathcal{G}_{od}) (\underline\beta^*)^2 \right]\;.
\end{split}
\end{equation}
Once again, one has to take a sufficiently involved model for the source to see the degeneracies, as a simple model like an elliptical Sérsic alone cannot present degeneracies with flexion parameters, only with the shear and convergence.

This transformation allows us to write the lens equation in the case of the minimal flexion model:
\begin{empheq}[box=\widefbox]{align}
    \Tilde{\underline\beta} &= (1-\kappa_{\rm LOS})\underline\theta - \gamma_{\rm LOS}\underline\theta^* - \frac{1}{2}\left(\mathcal{F}_{\rm LOS}^*\underline\theta^2 + 2\mathcal{F}_{\rm LOS}\underline\theta \, \underline\theta^*  + \mathcal{G}_{\rm LOS}^*(\underline\theta^*)^2  \right)  - \Tilde{\underline\alpha}(\underline\theta)  \label{lens eq final} \\
    \Tilde{\underline\alpha}(\underline\theta) &= \underline\alpha_{\rm eff}(\underline\theta) + \frac{1}{2} \left\{ \mathbb{F}^*_{\rm LOS}\underline\alpha^2_{\rm eff}(\underline\theta) + 2\mathbb{F}_{\rm LOS}\underline\alpha^*_{\rm eff}(\underline\theta) \underline\alpha_{\rm eff}(\underline\theta)  + \mathbb{G}_{\rm LOS}[\underline\alpha^*_{\rm eff}(\underline\theta)]^2 \right\} \;. \label{alpha 2tilde}
\end{empheq}

\noindent
Here, $\gamma_{\rm LOS} \equiv \gamma_{od} + \gamma_{os} - \gamma_{ds}$ (similarly for $\kappa_{\rm LOS}$), and $ \mathcal{F}_{\rm LOS} \equiv \mathcal{F}_{od} + \mathcal{F}_{os} - \prescript{(2)}{}{\mathcal{F}}_{ds}$ (and similarly for $\mathcal{G}_{\rm LOS}$) while $\mathbb{F}_{\rm LOS}$ is defined as 
\begin{equation}\label{def Flos}
     \mathbb{F}_{\rm LOS} \equiv \mathcal{F}_{od} + \prescript{(1)}{}{\F}_{ds}  - \prescript{(2)}{}{\mathcal{F}}_{ds},
\end{equation}
and the same goes for $\mathbb{G}_{\rm LOS}$. Notice that, starting with \cref{lens eq final}, we easily recover eq.~(3.17) from \cite{LOS_in_SGL} by taking the flexion parameters to zero, as $(1-\kappa_{\rm LOS})\underline\theta - \gamma_{\rm LOS}\underline\theta^*$ is the complex equivalent of $(1-\boldsymbol\Gamma_{\rm LOS})\boldsymbol\theta$. Looking at \cref{alpha 2tilde}, we see that we have been able to further reduce the number of flexion parameters. Here, the LOS flexion is described by only four complex parameters. We therefore call this the minimal flexion model, as it fully encodes all possible degeneracies in the model using the minimum number of parameters.

\begin{figure}[h]
    \centering
    \includegraphics[width=\textwidth]{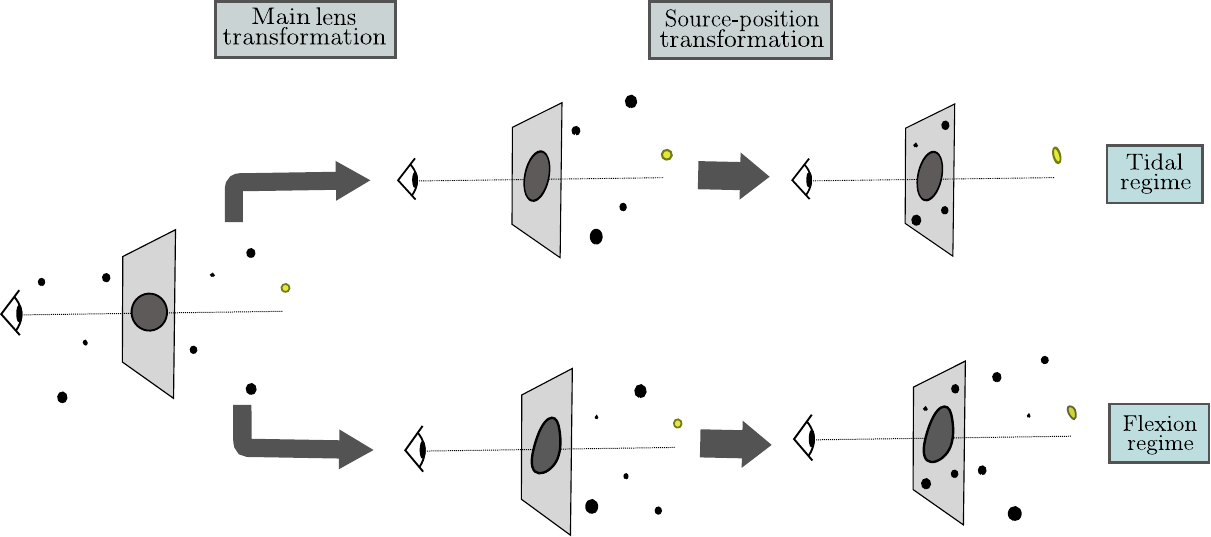}
    \caption{Sequence of transformations to build the minimal model, in the case of the tidal regime (top) and flexion regime (bottom). Black points represent LOS objects while the yellow points represent the source. In the first step we redefine the main lens, getting rid of the foreground. The background is also modified. In the second step, the source is distorted. In the tidal regime the shear and convergence are brought back in the main lens plane, but in the flexion regime, some background flexion terms remain. However, the number of parameters to describe the line of sight is reduced.}
    \label{fig: transformations}
\end{figure}

Let us now highlight several points. First, we can see that the two types of LOS flexion defined here ($\F_{\rm LOS}$ and $\mathbb{F}_{\rm LOS}$) have very different roles. The first one, just like the LOS shear, is mathematically equivalent to an external perturbation to the main lens within the same plane. On the other hand and quite remarkably, the second type of LOS flexion cannot be reabsorbed in the main lens plane; it is inherent to the line of sight. This is very different from the tidal approximation developed in \cite{LOS_in_SGL}, where one stops at the level of the shear. In the tidal approximation, starting from a main lens and a set of perturbers along the line of sight, we are able to bring all the perturbations into the main lens plane. This is impossible in the flexion regime: we will always be left with some LOS flexion. This means that $\mathbb{F}_{\rm LOS}$ and $\mathbb{G}_{\rm LOS}$ are not degenerate with the main lens or any substructure of the main lens, suggesting a way to distinguish LOS perturbers from haloes within the main lens plane. This possibility will be further investigated in future work. An illustration of the effect of the subsequent transformations is given in \cref{fig: transformations}.

We also note that other SPT may be considered. But it is impossible to reduce the number of flexion parameters below four, and the models they lead to will be equivalent to the minimal model defined in this work.

\subsection{Time delays} As described in the introduction, strong lensing time delays can be used to obtain constraints on the absolute distances in the Universe, and hence $H_0$. Our next step will be to compute time delays in the flexion regime. Note that the flexion corrections to time delays would be very small compared to the uncertainty involved by lens modelling and the aforementioned mass-sheet degeneracy, but we present them for the sake of completeness. 

Let us remind the reader that in the dominant lens formalism the time delay between two images A and B from the same source at $\underline{\beta}$ reads $\Delta t_{\rm AB} = T(\underline\theta_{\rm A}, \underline{\beta}) - T(\underline\theta_{\rm B}, \underline{\beta})$, where
\begin{equation} \label{Time delay start}
\begin{split}
    T(\underline\theta, \underline\beta) &= \frac{1}{2}\tau_{ds} |\underline\theta - \underline\alpha_{od} - \underline\beta|^2 - (1+z_d)\hat{\psi}_d[D_{od}(\underline\theta - \underline\alpha_{od})] \\
    &- \sum_{l<d}(1+z_l)\hat{\psi}_l(D_{ol}\underline\theta) - \sum_{l>d}(1+z_l)\hat{\psi}_l(D_{ol}(\underline\theta - \underline{\alpha}_{odl})]
\end{split}
\end{equation}

\noindent
at first order in $\epsilon$ (see \cite{LOS_in_SGL} for more details and a proof). As the only relevant quantity that one uses in time delays is the difference $\Delta t_{\rm AB}$, every component that does not depend on $\underline\theta$ in $T(\underline\theta, \underline\beta)$ will cancel when performing the subtraction, and is therefore irrelevant. The relevant part of the time delay is then
\begin{equation}\label{time delay final}
    \fbox{$\displaystyle \begin{split}
    T(\underline\theta)
    = \frac{\tau_{ds}}{2} \Re\Bigg\{ \underline\alpha_{\rm eff}^*(\underline\theta) \bigg[& \big\{ 1 + \kappa_{\rm LOS}(0) \big\}\underline\alpha_{\rm eff}(\underline\theta) + \gamma_{\rm LOS}(0)\underline\alpha_{\rm eff}^*(\underline\theta) \\
    &+ \frac{2}{3} \Big\{ \mathbb{F}_{\rm LOS}^*\underline\alpha_{\rm eff}^2(\underline\theta) + 2 \mathbb{F}_{\rm LOS}|\underline\alpha_{\rm eff}(\underline\theta)|^2 + \mathbb{G}_{\rm LOS}\underline\alpha_{\rm eff}^{*2}(\underline\theta)  \Big\} \\
    &+ \Big\{ \F_{\rm LOS}^*\T^2 + 2 \F_{\rm LOS}|\T|^2 + \G_{\rm LOS}\T^{*2}  \Big\} \bigg]\Bigg\} \\
    -\frac{\tau_{ds}}{2} \Re\Bigg\{ \mathcal{F}_{\rm LOS}\T & (\T^*)^2 + \frac{1}{3} \mathcal{G}_{\rm LOS}(\T^*)^3 \Bigg\}- \tau_{ds}\psi_{\rm eff}(\underline\theta) \;.
    \end{split}$}
\end{equation}
In this equation, $\Re$ stands for the real part, and $\tau_{ds}$ is the usual time delay distance scale defined as $\tau_{ds} \equiv (1+z_d)D_{od}D_{os}/D_{ds}$. Once again, when taking the flexion parameters to zero we recover the result of \cite{LOS_in_SGL}, as $\Re(\underline a^* \underline b) = \boldsymbol a \cdot \boldsymbol{b}$.

\input{Mock_images}

\section{Conclusion} \label{sec:conclusion}

In this paper, we have built a model to account for LOS effects going beyond the tidal regime parameterised solely by convergence and shear. Using this new tool, we have shown that the measurement of the LOS shear from mock strong lensing images can be jeopardised if one neglects flexion effects.

More precisely, in \cref{sec:LOStheory}, we have reviewed the LOS formalism and the dominant lens approximation first developed in \cite{LOS_in_SGL}. In this model, we consider that a dominant lens is the main contributor to the deflection of light rays, whilst inhomogeneities along the line of sight are treated as perturbers. This model allows us to correctly account for LOS effects while being simpler to handle than the fully general multi-plane lensing formalism. We then presented the form the lens equation takes if one goes beyond shear, expanding the projected gravitational potential of the perturbers at third order, which gives rise to flexion terms.

Starting with this framework, we built a so-called minimal model for flexion in \cref{Section : Minimal flexion model}, generalising what was done in \cite{LOS_in_SGL} at the level of the shear. This model allows one to treat LOS effects up to flexion with the minimum possible number of parameters, avoiding degeneracies that could enlarge the error bar of the measured parameters. We have shown that, unlike the case of the convergence or the shear, no source-position transformation allows one to absorb all of the flexion in the main lens plane. Namely, the flexion parameters which we called $\FF$ and $\GG$, are markers of the LOS haloes. This suggests a way to distinguish LOS perturbers from perturbers within the main lens plane. This possibility will be further investigated in future work.

In \cref{Section: LOS flexion}, we addressed the measurability of the LOS flexion from mock strong lensing images. To put a realistic value of the flexion in our simulations, we computed the expected variance of cosmic flexion in $\Lambda$CDM. As flexion is highly sensitive to small scales, and as the matter power spectrum is not known at those scales, we needed to fill in the gaps. However, a simple power law extrapolation of the aforementioned power spectrum results in the divergence of the variance of the flexion. To solve this problem, we instead computed the variance of the averaged flexion on the scale of the image, as we assume the flexion to be homogeneous across the image in the flexion regime. Adding the flexion model to \lenstronomy, we have been able to show the superiority of the minimal model described by \cref{alpha 2tilde} compared to the full model encoded in \cref{alpha} when fitting mock strong lensing images.

Finally, we tested the measurability of the LOS shear in presence of LOS flexion in \cref{Sec: LOS shear measure}. We found that neglecting flexion using the minimal LOS model used in \cite{Hogg_2023} can lead to bias in shear measurements. Instead, the minimal flexion model can provide an accurate measure of the shear at the cost of worsened precision. This means that the conclusions of Hogg et al. \cite{Hogg_2023} regarding flexion are somewhat optimistic, as in the presence of LOS flexion the error bars on the shear will be enlarged. Furthermore, the bias appears even for relatively small values of the LOS flexion (of order of a few $10^{-3}\;\rm arcsec^{-1}$), in which case we cannot distinguish flexion from the noise. This likely means that with current imaging capabilities we cannot measure flexion with a single image. However, as noise in different lensing images should not be correlated, it might still be possible to use flexion to probe small scale properties of dark matter using correlation functions.

This work highlights the importance of having a proper estimate of weak lensing flexion in the Universe, since the precision of LOS shear measurements depend on its magnitude in a given image. But, as explained earlier, we currently cannot compute the expected variance of the cosmic flexion without resorting to some averaging, limiting our theoretical predictions. Besides, current $N$-body simulations are not helpful either, as their mass resolutions are typically too coarse to include low-mass haloes, which play the dominant role in the production of flexion in a strong lensing image.  Moreover, as the measurement of flexion is very challenging, we do not have a robust value coming from current observational data. Addressing the problem of the flexion magnitude is left for future work. 


\section*{Acknowledgements}

NBH is supported by a postdoctoral position funded by IN2P3. PF acknowledges support from the French \emph{Agence Nationale de la Recherche} through the ELROND project (ANR-23-CE31-0002). The authors thank Daniel Johnson for some valuable input on an advanced version of this paper.

\section*{\href{https://www.elsevier.com/authors/policies-and-guidelines/credit-author-statement}{CRedIT} authorship contribution statement}

\noindent \textbf{Théo Duboscq:} Conceptualisation, Methodology, Software, Investigation, Writing -- Original draft, Visualisation.
\textbf{Natalie B. Hogg:} Software, Writing -- Review \& Editing.
\textbf{Pierre Fleury:} Conceptualisation, Writing -- Review \& Editing.
\textbf{Julien Larena:} Conceptualisation, Methodology, Validation, Writing -- Review \& Editing, Supervision.

\bibliographystyle{JHEP.bst}
\bibliography{main.bib}

\input{Annexe}

\end{document}

%% file: packages.tex
\usepackage{jcappub}

\usepackage{mathrsfs}
\usepackage{bbm}
\usepackage{bm}
\usepackage{dsfont}
\usepackage{tensor}
\usepackage{xspace}
\usepackage{lmodern}
\usepackage{wasysym}
\usepackage{microtype}
\usepackage{empheq}
\usepackage{ifthen}
\usepackage{amsmath}
\usepackage{wrapfig}
\usepackage{cleveref}
\crefname{section}{sec.}{secs.}
\Crefname{section}{Section}{Sections}
\crefname{subsection}{subsec.}{subsecs.}
\Crefname{subsection}{Subsection}{Subsections}
\crefname{subsubsection}{subsubsec.}{subsubsecs.}
\Crefname{subsubsection}{Subsubsection}{Subsubsections}
\usepackage{xcolor}
\usepackage{import}
\usepackage{transparent}

%% file: macros.tex
\newcommand\define{\equiv}

\renewcommand\Re{\mathrm{Re}}

\newcommand\h[1]{^{\mathrm{#1}}}

\newcommand{\delimiters}[4][]{
\ifthenelse{ \equal{#1}{1} }{  #2 #3 #4  }
					{ \ifthenelse{\equal{#1}{2}}{ \big#2 #3 \big#4 }
						{ \ifthenelse{\equal{#1}{3}}{ \Big#2 #3 \Big#4 }
							{ \ifthenelse{\equal{#1}{4}}{ \bigg#2 #3 \bigg#4 }
								{ \ifthenelse{\equal{#1}{5}}{ \Bigg#2 #3 \Bigg#4 }
									{ \left#2 #3 \right#4 }
								}
							}
						}
					}
													}

\definecolor{blue4}{RGB}{0,0,143}
\definecolor{red4}{RGB}{143,0,0}
\definecolor{orange}{RGB}{255,128,0}
\definecolor{darkcyan}{RGB}{0,128,128}
\definecolor{olive}{RGB}{0,128,0}
\definecolor{purple}{RGB}{128,0,128}
\definecolor{cyan2}{RGB}{0,255,255}
\definecolor{fushia}{RGB}{255,0,255}
\definecolor{mygray}{gray}{0.5}
\definecolor{lightgray}{gray}{0.85}


%% file: Mock_images.tex
\section{Measuring LOS flexion in strong lensing images} \label{Section: LOS flexion}

At this stage, a natural question to ask is whether the LOS flexion can be measured from strong-lensing images, as it could give us extra information about the matter distribution around the line of sight, especially since flexion is sensitive to the small scale distribution of matter \cite{Bacon_2006}. Of course, this will depend on its magnitude; if the LOS flexion is too small, it will have a correspondingly small effect on a given lensing image, and we might not be able to measure it. To assess the measurability of the LOS flexion, we generate strong lensing images using \lenstronomy, limiting our analysis to Einstein rings, and perform Markov chain Monte Carlo~(MCMC) parameter inference to recover the parameters that were used to create the image in the first place, similarly to what was done in \cite{Hogg_2023}. To do this, we use the LOS flexion formalism we have implemented in \lenstronomy mentioned in \cref{sec:LOStheory}. The use of the LOS flexion in \lenstronomy is similar to that of the LOS shear, with the user simply needing to call the relevant lens model profile and define the corresponding LOS parameters.

\subsection{The expected value of cosmic flexion} \label{Section: flexion value}

In order to create the mock images with realistic values for the flexion parameters, we computed the expected cosmic flexion variance predicted by the spatially flat $\Lambda$CDM cosmological model described by the \textit{Planck} 2018 data \cite{Planck2018}, with $H_0 = 67.4\; \rm km s^{-1} Mpc^{-1}$ and $\Omega_{\rm m,0} = 0.315$. To do so, we first need the expression of the cosmic flexion in terms of the matter density contrast $\delta$. We define a line of sight and an orthonormal basis on the flat sky (which defines a complex plane). The detailed calculations are given in \cref{Appendix: cosmic calc}. We get the following expressions:
\begin{align}
    \mathcal{F}(\T) &= \frac{3}{2}\Omega_{\rm m,0} H_0^2 \int_0^{\chi_s} {\rm d}\chi \; (1+z) W(\chi) \Delta[\eta_0-\chi, \chi, f_K(\chi)\T] \;, \label{Final cosmic F}\\
    \mathcal{G}(\T) &= - 3 \Omega_{\rm m,0} H_0^2 \int_0^{\chi_s} {\rm d}\chi \; (1+z) W(\chi) \int_{\mathbb{R}^2} \frac{\mathrm{d}^2\boldsymbol{\zeta}}{\pi \zeta^3} {\rm e^{3i\phi}}\delta[\eta_0-\chi, \chi, \underline{\zeta}+f_K(\chi)\T] \\
    &= - \frac{3}{2} \Omega_{\rm m,0} H_0^2 \int_0^{\chi_s} {\rm d}\chi \; (1+z) W(\chi) \int_{\mathbb{R}^2} \frac{\mathrm{d}^2\boldsymbol{\zeta}}{\pi \zeta^2} {\rm e^{2i\phi}}\Delta[\eta_0-\chi, \chi, \underline{\zeta}+f_K(\chi)\T]\;. 
\end{align}
In the above, $\Omega_{\rm m, 0}$ is the dimensionless matter density at redshift zero, $H_0$ is the Hubble parameter at redshift zero, $\delta(\eta, \chi, \underline\zeta)$ is the matter density contrast at conformal time $\eta$, comoving distance $\chi$ and transverse position $\underline\zeta = \zeta \rm e^{i\phi}$ from the optical axis. The parameter $\Delta \define \partial \delta/\partial \underline\zeta^*$ is the complex equivalent of half of the gradient of the matter density contrast across a plane orthogonal to the line of sight. Then, $\F$ can be any of the type-$\F$ flexion parameters ($\F_{od}$, $\F_{os}$, $^{(1)}\F_{ds}$, $^{(2)}\F_{ds}$, $\F_{\rm LOS}$ or $\FF$), and $W$ is the corresponding weight function. The same goes for $\G$. We give the expression of all of these weight functions in \cref{Appendix: cosmic calc}. Notice that the weight functions for a given $\F$ and the corresponding $\G$ are the same, which reflects the fact that the same geometric factors enter in the respective definitions.

With the above in hand, we can now define the autocorrelation functions of the flexion in order to obtain its variance. The previous formulae involve LOS integrals of the matter density contrast $\delta$, so one can foresee that the autocorrelation functions of those quantities will be expressed in terms of the matter power spectrum.
We consider a direction $\underline\vartheta$ in the sky, in which one can observe several strong lensing systems, for which the LOS flexion parameters could be measured. Apart from the position $\underline\vartheta$, the flexion depends on the comoving position $\chi_d$ of the main deflector of the system, and the comoving position $\chi_s$ of the source. Let us call $\boldsymbol\Pi \equiv (\chi_d, \chi_s)$ the set of these parameters, and $p$ their joint probability density function. Then we define the effective LOS flexion as 
\begin{equation} \label{F eff}
    \mathcal{F}_{\rm eff}(\underline{\vartheta}) \equiv \int p(\boldsymbol{\Pi}) \mathcal{F}(\underline{\vartheta}, \chi_d, \chi_s)  {\rm d}^2\boldsymbol\Pi \;.
\end{equation}
The effective LOS type-$\mathcal{G}$ flexion is defined in the same way. The flexion variance is then defined as:
\begin{equation}
    \sigma^2_{\mathcal{F}} = \langle \mathcal{F}_{\rm eff}(\underline{\vartheta}) \mathcal{F}_{\rm eff}^*(\underline{\vartheta}) \rangle \;,
\end{equation}
where we average over $\underline\vartheta$, and similarly for $\G$. Here we will skip the calculation. The interested reader can refer to \cref{Appendix: variance1} for more details. The final result reads:
\begin{equation} \label{sigma1}
    \sigma_\F^2 = \sigma_\G^2 = \int_0^{\infty} \frac{\ell^3}{2\pi} P_{\rm F}(\ell) \rm d\ell \;,
\end{equation}
with
\begin{align}
    P_{\rm F}(\ell) &= \frac{1}{4}\left(\frac{3}{2}\Omega_{\rm m,0} H_0^2\right)^2  \int_{0}^{\chi_s} {\rm d}\chi \; (1+z)^2 q_{\rm F}(\chi)^2 P_{\delta}\left(\eta_0 - \chi, \frac{\ell}{f_K(\chi)} \right) \;, \\
    q_{\rm F}(\chi) &= \int {\rm d}^2\boldsymbol\Pi \; p(\boldsymbol{\Pi}) \frac{W(\chi, \boldsymbol\Pi)}{f_K^2(\chi)} \;.
\end{align}
Our results are consistent with the ones first obtained in \cite{Bacon_2006, Arena_2022} in the context of weak lensing. As a matter of fact, the flexion formalism was naturally first introduced in the context of cosmic weak lensing in \cite{Bacon_2006}. The only difference lies in the specific shape of the weight function $W$; the weak lensing scenario corresponds to $W_{os}$ in the present paper. \footnote{Beware, our definition of $\F$ and $\G$ differs from the ones of \cite{Bacon_2006, Arena_2022} by a factor $1/2$, which explains the factor 1/4 in our flexion power spectrum.}
It turns out, however, that according to the above definitions the variance of flexion diverges. Since the flexion is one derivative higher than the shear, it strongly depends on small scales. This results in an additional $\ell^2$ in the expression of the variance.\footnote{The shear variance has a similar form to the flexion variance, but with a different kernel $q(\chi)$ and $\ell$ instead of $\ell^3$.} This means that flexion is highly sensitive to high $\ell$, and if we extrapolate $P_{\delta}$ to high $\ell$ using a power law, the variance diverges. Thus, the value we obtain strongly depends on the upper bound we set in the numerical integral. 

\begin{figure}
    \centering
    \includegraphics[width=0.6\textwidth]{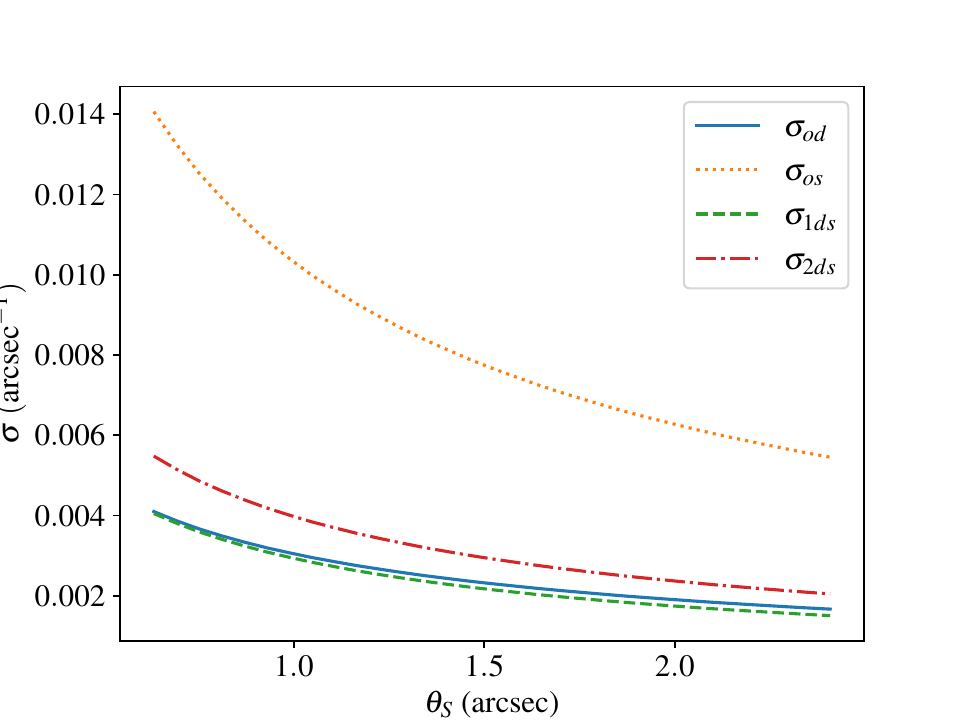}
    \caption{
       Standard deviation of the modulus of the cosmic flexion, depending on the scale $\theta_S$ on which we average the flexion. Since we assume that flexion is homogeneous on the scale of the image, we want a smoothing scale $\theta_S$ which corresponds roughly to the scale of an image. This result is the same for both flexion types, see \cref{eq: sigma with bessel}.
    } 
    \label{Fig: Sigmas F}
\end{figure}

To solve this problem, we will slightly change the definition of the quantity we are computing. In the flexion regime we assumed that flexion is homogeneous on the scale of the image, which we will call $\theta_S$. In this context, this means that for a given strong lensing system located at $\underline\vartheta$, we have
\begin{equation}
    \mathcal{F}_{\rm eff}(\underline{\vartheta}) = \frac{1}{\pi\theta_S^2} \int_S \F_{\rm eff}(\underline{s}) {\rm d}^2\boldsymbol{s} = \frac{1}{\pi\theta_S^2} \int_S {\rm d}^2\boldsymbol{s} \int {\rm d}^2\boldsymbol\Pi \; p(\boldsymbol{\Pi}) \mathcal{F}(\underline{s}, \chi_d, \chi_s)  \;,
\end{equation}

\noindent
where $S$ is the surface defined by the disk on the flat sky centred on $\underline\vartheta$ and of radius $\theta_S$. With this definition we can compute the variance again. The result is (see \cref{Appendix: variance2} for details):
\begin{equation} \label{eq: sigma with bessel}
    \sigma_\F^2 = \sigma_\G^2 = \int_0^{\infty} \frac{\ell^3}{2\pi} P_{\rm F}(\ell) \left[ \frac{2 J_1(\ell\theta_S)}{\ell\theta_S} \right]^2 \rm d\ell
\end{equation}
where $J_1$ is a Bessel function of the first kind. We see that this looks like \cref{sigma1} with an additional factor $[2J_1(\ell\theta_S)]^2/(\ell\theta_S)^2$, which makes the integral converge. Of course, the value we will find for the variance will now depend on the smoothing scale $\theta_S$ on which we average the flexion. This dependence is shown in \cref{Fig: Sigmas F}. 
This gives us a guideline for our mock images, as it gives us the expected scale the flexion should have. As previously stated, the scale $\theta_S$ we average over should correspond to the scale of the lensed images. We choose this scale to be the effective Einstein radius of the main lens. Of course, when taking $\theta_S$ to zero the variance of the flexion diverges, as it is equivalent to not taking the average anymore, hence recovering~\cref{sigma1}. Looking at those curves, we see that the variance of the $os$ flexion is greater than the others. This is due to the kernel $W_{os}(\chi, \boldsymbol{\Pi})$ involved in the expression of the variance (\cref{Fig: Weight functions} in \cref{Appendix: cosmic calc} shows a comparison of the different flexion kernels). Because of this, one might be tempted to only keep the $os$ flexion terms and neglect the others. We should refrain from doing so; first because there is only a factor 3 between their standard deviation, second because this would mean taking $\FF$ and $\GG$ to 0, losing a key term of the lens equation. We therefore do not make such an approximation, and keep all the terms for the subsequent analysis.
Finally, let us point out that this method is not the only renormalization solution. The authors of \cite{Arena_2022} propose, based on \cite{Widrow2009}, an alternative approach. They compute the power spectrum up to a large $k_{max}$ and then impose a steeper slope of the power spectrum above this threshold, resulting in the convergence of the integral. With this method, they find a variance quite similar to ours; looking at fig. 3 of \cite{Arena_2022} we find $\sigma_\F = \sigma_\G \approx 7\times 10^{-3} \, \rm arcsec^{-1}$, which is close to our $\sigma_{\F_{os}}=\sigma_{\G_{os}}$ for the $\theta_S$ of a typical strong lensing image ($\theta_S\sim 1$ arcsec).

\subsection{Advantages of the minimal flexion model} \label{Section: min model adv}

As we have seen in \cref{Section : Minimal flexion model}, we deliberately designed the minimal flexion model to avoid the degeneracies present in the flexion regime. To demonstrate the superiority of the minimal model over the full model, we generate a strong lensing image using \lenstronomy. As stated before, some flexion parameters are only degenerate with a complex enough main lens, e.g. $\F_{od}$ is not degenerate with any of the parameters of a SIS, because such a model is too simplistic. Therefore, to fully illustrate the advantage of the minimal model, we take a complex model, which aims to be more realistic. 

\begin{figure}
    \centering
    \includegraphics[width=0.35\textwidth]{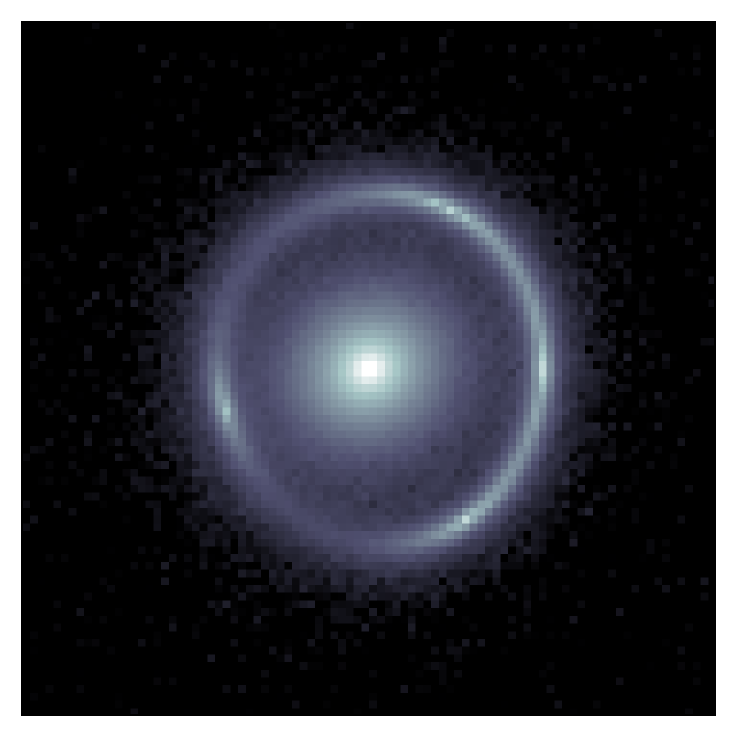}
    \caption{
       The image generated with \lenstronomy using the model described in \cref{Section: min model adv}. The parameters are taken at random according to the procedure of \cref{Section: min model adv} and of appendix B of \cite{Hogg_2023}.
    } 
    \label{Fig: ring}
\end{figure}

We use a main lens composed of two components, a baryonic one modelled by an elliptical Sérsic, and a dark matter component modelled by an elliptical Navarro–Frenk–White (NFW) profile \cite{Navarro:1995iw}. We allow for an offset between the centres of the two components. In both profiles, the ellipticity is injected at the level of the potential and not the convergence. Then, the lens light is modelled by an elliptical Sérsic, whose parameters are the same as the baryonic component of the main lens, so that the light profile follows the potential. The source light is then also modelled using an elliptical Sérsic model. We then add the LOS shear and flexion using the lens model we implemented in \lenstronomy. For each flexion (\textit{od}, \textit{os} and the two \textit{ds} terms), we draw the modulus square from a uniform distribution $|\F|^2 \sim \mathcal{U}(0, \F_{\rm max}^2)$, with $\F_{\rm max}^2$ being twice the cosmic flexion variance of the given component computed in \ref{Section: flexion value}. Then, we compute the two flexion components according to
\begin{align}
    \F_1 &= \sqrt{|\F|^2} \cos{\phi} \\
    \F_2 &= \sqrt{|\F|^2} \sin{\phi} \;,
\end{align}
with $\phi \sim \mathcal{U}(0, 2\pi)$. We do the same for type-$\G$ flexion. All other parameters used to generate the image are taken at random using the exact same procedure as the one used in sec. 3 of \cite{Hogg_2023}. The interested reader can refer to appendix B of \cite{Hogg_2023} for more details. Finally, we simulate the noise and point spread function according to the Hubble Space Telescope Wide Field Camera 3 F160W noise settings in \lenstronomy \cite{Windhorst_2011}. This gives the lensing image shown in \cref{Fig: ring}.

We then fit a model to the image using an MCMC parameter inference method, using the ensemble
slice sampling algorithm implemented in the Python package \texttt{zeus}\footnote{\url{https://github.com/minaskar/zeus}.} \cite{Karamanis:2020zss, Karamanis:2021tsx}. However, as stated in \cref{Section : Minimal flexion model}, one needs a sufficiently involved source model to perform the source transformation \cref{SPT}. Namely, an elliptical Sérsic does not become another elliptical Sérsic under this transformation. This is quite different from the case of the tidal regime, because in the latter the shear and the convergence are degenerate with the ellipticity of the Sérsic and the angular half-light radius $R_{\textrm{Sérsic}}$ respectively. However, in the flexion regime, flexion is not degenerate with any of the Sérsic parameters, meaning that if we fit the image using the minimal model and a Sérsic for the source light, it will fail to properly recover the flexion parameters. This is because the minimal model assumes we perform the aforementioned SPT. 

To alleviate this problem, we need to fit the source light with a sufficiently complex model. As we are not interested in the shape of the source, we can fit the source with any model we want, as long as this model can reproduce the shape of a Sérsic transformed by the SPT given in \cref{SPT}. To this end there are several options, such as using a pixellised source model \cite{Warren2003, Vegetti2009}
or a basis function regression, e.g. shapelets \cite{Refregier1_2003, Refregier2_2003}. However, these methods are more computationally expensive than using an analytic model for the source, which slows down an MCMC parameter inference. To ensure timely convergence, we constructed an analytic model with an accompanying profile in \lenstronomy called \texttt{SERSIC\_FLEXION}.\footnote{This light profile is also in the \texttt{los-flexion} branch of \texttt{lenstronomy} on Github.} This model consists of a Sérsic with the addition of flexion,
\begin{equation} \label{Sersic flexion}
    I(\T) = I_{\textrm{Sérsic}}\Big[(1+a)\T + b\T^* + c^*\T^2 + 2c|\T|^2 + d\T^{*2}\Big] \;,
\end{equation}
with $I_{\textrm{Sérsic}}$ being the profile of an elliptical Sérsic. This means that if a source is described by a Sérsic profile, then after the transformation \cref{SPT} it is described by \cref{Sersic flexion} with $a=\kappa_{ds}-\kappa_{od}$, $b=\gamma_{ds}-\gamma_{od}$, $c=\prescript{(2)}{}{\F}_{ds} - \F_{od}$ and $d=\prescript{(2)}{}{\G}_{ds} - \G_{od}$. Note that as the shear is degenerate with the Sérsic ellipticity, and the convergence is degenerate with the half-light radius, so we can keep $a$ and $b$ fixed to zero, instead of sampling them in our MCMC analysis.

\begin{figure}
    \centering
    \begin{minipage}{0.49\textwidth}
        \includegraphics[width=\textwidth]{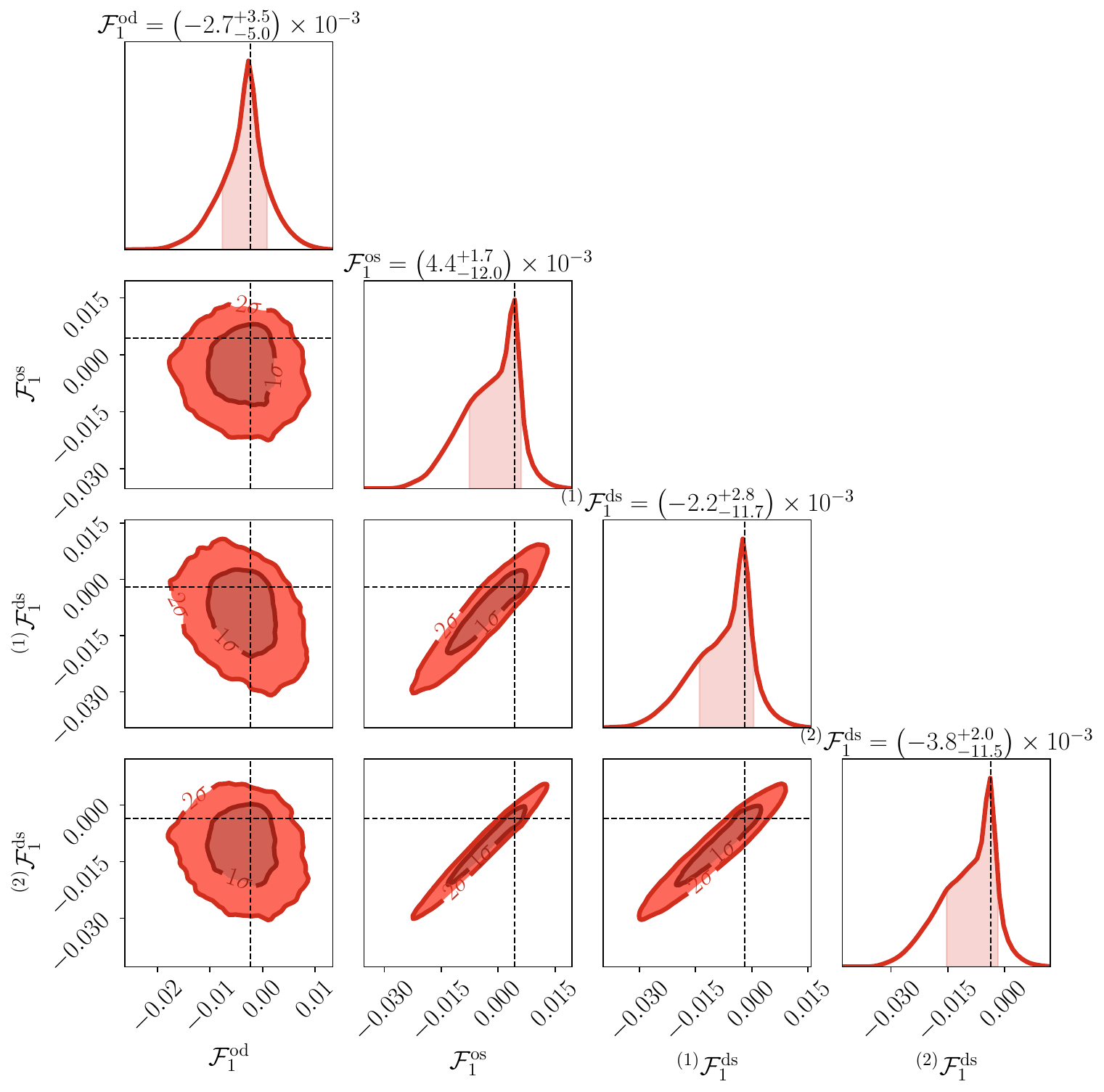}
    \end{minipage} \hfill
    \begin{minipage}{0.49\textwidth}
        \includegraphics[width=\textwidth]{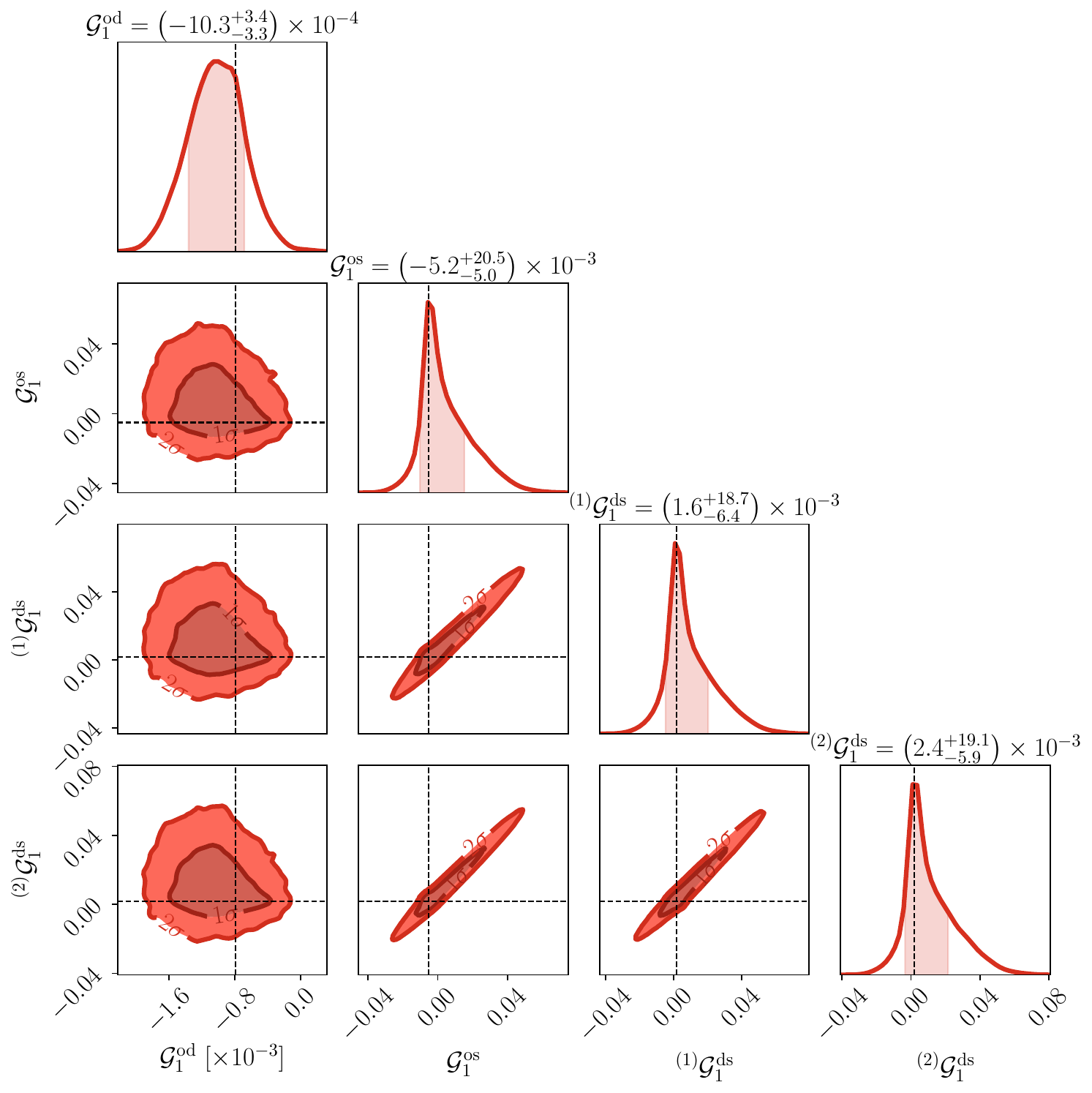}
    \end{minipage}
    \caption{
       \textbf{Inference with the full flexion model}, fitting the image in \cref{Fig: ring}. \textit{Left panel:} one- and two-dimensional marginalised posterior distributions of the various type-$\F$ flexion. \textit{Right panel:} one- and two-dimensional marginalised posterior distributions of the various type-$\G$ flexion. The dotted lines represent the input value of each parameter, i.e. the one used to generate the image. For readability we only show the real part of each parameter, but the imaginary part exhibits similar contours.
    } 
    \label{Fig: contours LOSF full}
\end{figure}

In \cref{Fig: contours LOSF full}, we show the one- and two-dimensional marginalised posterior distributions of the various flexion parameters involved in the full flexion model, i.e. the one defined by \cref{alpha}. Note that all the contours shown in this paper are obtained with the \texttt{chainconsumer} package~\cite{Hinton2016}.\footnote{\url{https://github.com/Samreay/ChainConsumer}.} Note that here we do not show the two-dimensional posteriors of the type-$\F$ and type-$\G$ flexion together, but doing this one can see that none of the type-$\F$ and type-$\G$ flexion parameters are degenerate with each other. This is unsurprising since their effect is highly different: type-$\F$ flexion is a spin-1 quantity while type-$\G$ flexion is a spin-3 quantity. However, it is evident that there are strong degeneracies between different flexion parameters of the same type ($\F$ or $\G$). Furthermore, the posteriors are strongly non-Gaussian and the error bars are large, often four to ten times larger than the flexion values themselves. 

\begin{figure}
    \centering
    \includegraphics[width=0.49\textwidth]{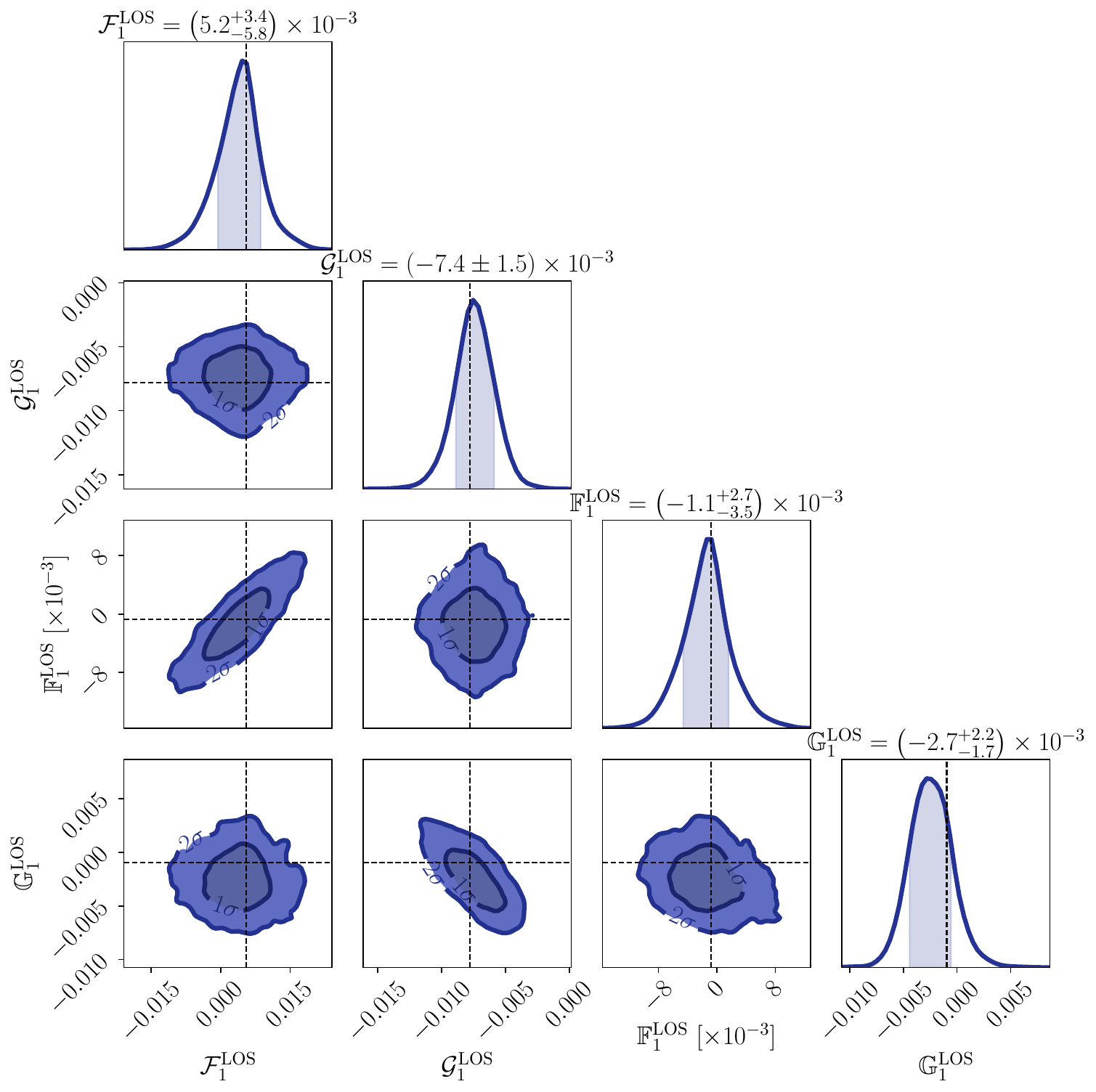}
    \caption{
       \textbf{Inference with the minimal flexion model}, fitting the image in \cref{Fig: ring}: one- and two-dimensional marginalised posterior distributions of the $\F_{\rm LOS}$, $\G_{\rm LOS}$, $\FF$ and $\GG$. The dotted lines represent the input value of each parameter, i.e. the one used to generate the image. For readability we only show the real part of each parameter, but the imaginary part exhibits similar contours.
    } 
    \label{Fig: contours LOSF min}
\end{figure}

We can contrast this with the posteriors we get when fitting the same image using the minimal flexion model, defined in \cref{alpha 2tilde} and shown in \cref{Fig: contours LOSF min}. The posteriors obtained on the flexion parameters using the minimal model are more symmetric than the results obtained using the full model. The error bars are also globally smaller, by about $3 \times 10^{-3} \rm arcsec^{-1}$ on average. In this example, we can detect a non-zero $\G_{\rm LOS}$ at the 5$\sigma$ level. Nevertheless, these results also highlight an obvious difficulty with the potential measurability of flexion: since the expected cosmic signal is very small, current imaging data and inference procedures may not allow us to distinguish it from the noise.

\section{Measuring LOS shear in the presence of flexion}
\label{Sec: LOS shear measure}

As proposed in refs. \cite{LOS_in_SGL} and \cite{Hogg_2023}, LOS shear could be a cosmological observable if one could measure it with enough precision. It was shown in the latter that, in an ideal setup where flexion is negligible, the minimal LOS model could lead to measurements of the LOS shear with sufficient precision for cosmological constraints. However, the same work hinted that the precision of such a measurement will be worsened when flexion is not negligible (see subsec.~4.3 of \cite{Hogg_2023}). This suggests that flexion may have to be accounted for in a realistic situation. However, using the minimal flexion model built in \cref{Section : Minimal flexion model} leads to the addition of several parameters to the lens model. The goal of this section is therefore to investigate the effect of adding flexion to images for the subsequent reconstruction, and to see if the minimal flexion model performs better than the minimal LOS model. For this task, we again use mock images created with \lenstronomy.

\begin{figure}
    \centering
    \includegraphics[width=0.8\textwidth]{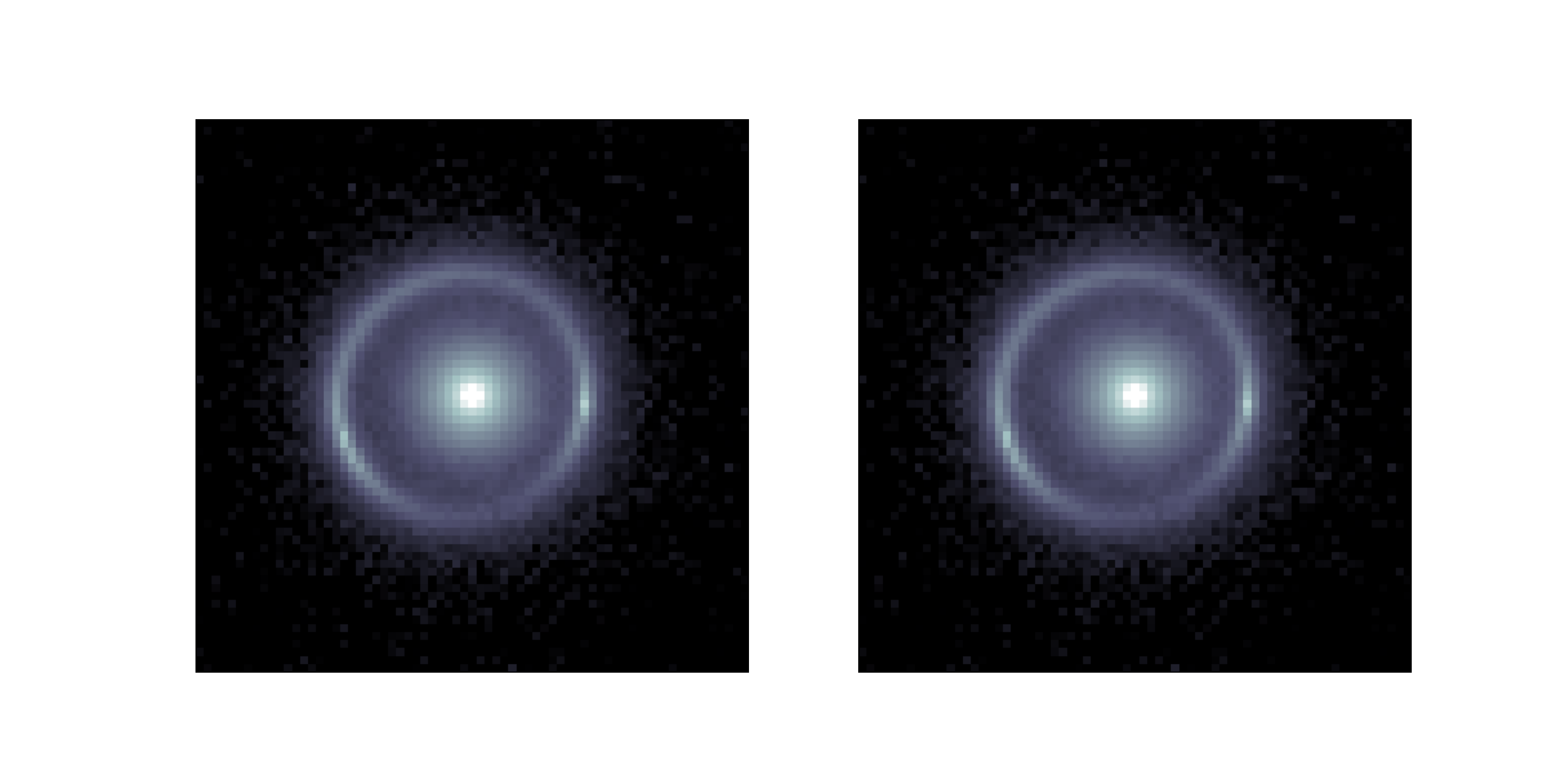}
    \caption{
       The images generated with the model described in \cref{Sec: LOS shear measure}. \textit{Left panel:} The image generated with flexion parameters taken accordingly to the value of the expected cosmic flexion computed in \cref{Section: flexion value}. \textit{Right panel:} The image generated with the flexion parameters rescaled by a factor 0.3.
    } 
    \label{Fig: rings}
\end{figure}

To generate the simulated images, we designed the following procedure. We use an EPL profile to create a simple main lens, with an Einstein radius $\theta_{\rm E} = 1.2 ''$, a negative power-law slope $\gamma=2.6$, and an ellipticity taken at random. More precisely, we compute the ellipticity with the following expressions:
\begin{align}
    e_1 = \frac{1-q}{1+q} \, \cos{2\phi} \;, \\
    e_2 = \frac{1-q}{1+q} \, \sin{2\phi} \;, 
\end{align}
with $q$ the aspect ratio and $\phi$ the orientation angle of the ellipse. The parameters are drawn at random from the uniform distributions $q\sim \mathcal{U}(0.9, 1.0)$ and $\phi\sim \mathcal{U}(0, 2\pi)$. The centre of the profile is fixed at the centre of the image coordinates. We add lens light and source light the same way as described in the previous section, following the procedure presented in appendix~B of \cite{Hogg_2023}. We also add LOS shear and flexion as in the previous section. The resulting image is shown in the left panel of \cref{Fig: rings}. 

Then, to see the impact of the value of the flexion, we generate an image with the exact same parameters but with the flexion parameters rescaled by a factor 0.3. Doing so, most flexion parameters are $\lesssim 1 \times 10^{-3}\; \rm arcsec^{-1}$. The flexion in this case can be considered negligible. This image is shown in the right panel of \cref{Fig: rings}. We then fit each image with two different models: on the one hand the minimal LOS model, which only takes into account LOS shear (see \cite{LOS_in_SGL, Hogg_2023} for details), and on the other hand the minimal flexion model developed in \cref{Section : Minimal flexion model}, which adds several parameters by taking flexion into account. 

\begin{figure}
    \centering
    \includegraphics[width=0.8\textwidth]{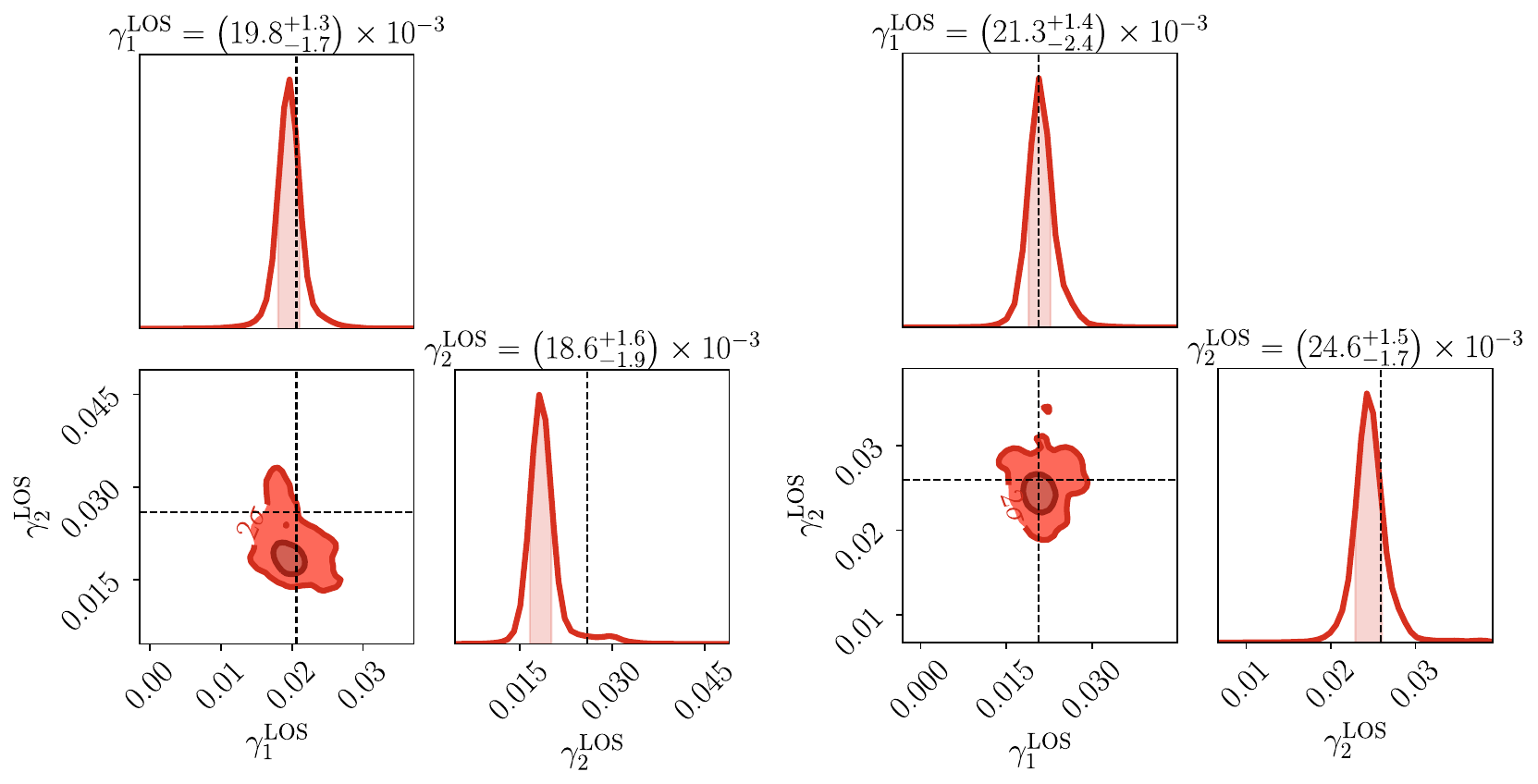}
    \caption{
       \textbf{Inference with the minimal LOS model}, fitting the images in \cref{Fig: rings}, where we show the one- and two-dimensional marginalised posterior distributions of the LOS shear. The dotted lines represent the input value of each parameter, i.e. the one used to generate the image.
    } 
    \label{Fig: contours LOS}
\end{figure}

The one- and two-dimensional marginalised posterior distributions of the shear parameters using the minimal LOS model are shown in \cref{Fig: contours LOS}. Looking at them we see that the minimal LOS model is not sufficient to fit the image when flexion is not negligible. From the left panel in this figure, it is clear that in presence of flexion the MCMC fails to recover the fiducial value of the shear: the inferred value is more that 4$\sigma$ away from the truth for $\gamma_2^{\rm LOS}$. This means that using the minimal LOS model could lead to a significant bias in the reconstruction of the LOS shear if flexion is not negligible. This holds true even for relatively small values of the flexion; for example, the flexion used to generate the image in the left panel is a few $10^{-3} \;\rm arcsec^{-1}$ for most flexion parameters, less than $10^{-2} \;\rm arcsec^{-1}$ for the largest one; see \cref{tab: los params} for details.
\begin{table}
  \centering
  \begin{tabular}{SlSl|SlSl}
    \hline
    \hline
    \multicolumn{2}{c}{Expected flexion}&\multicolumn{2}{|c}{Rescaled flexion} \\
    \hline
    Parameter  & Value & Parameter  & Value \\
\hline
$\gamma_1\h{LOS}$   & $2.1\times10^{-2}$ & $\gamma_1\h{LOS}$   & $2.1\times10^{-2}$\\
$\gamma_2\h{LOS}$   & $2.6\times10^{-2}$ & $\gamma_2\h{LOS}$   & $2.6\times10^{-2}$ \\
$\F_1\h{LOS}$     & $-2.4\times10^{-3} \;\rm arcsec^{-1}$ & $\F_1\h{LOS}$     & $-7.2\times10^{-4} \;\rm arcsec^{-1}$ \\
$\F_2\h{LOS}$     & $8.5\times10^{-3} \;\rm arcsec^{-1}$ & $\F_2\h{LOS}$     & $2.5\times10^{-3} \;\rm arcsec^{-1}$ \\
$\G_1\h{LOS}$     & $-1.5\times10^{-3} \;\rm arcsec^{-1}$ & $\G_1\h{LOS}$     & $-4.5\times10^{-4} \;\rm arcsec^{-1}$ \\
$\G_2\h{LOS}$     & $-2.5\times10^{-3} \;\rm arcsec^{-1}$ & $\G_2\h{LOS}$     & $-7.5\times10^{-4} \;\rm arcsec^{-1}$ \\
$\mathbb{F}_1\h{LOS}$     & $2.3\times10^{-3} \;\rm arcsec^{-1}$ & $\mathbb{F}_1\h{LOS}$     & $6.9\times10^{-4} \;\rm arcsec^{-1}$\\
$\mathbb{F}_2\h{LOS}$     & $1.8\times10^{-3} \;\rm arcsec^{-1}$ & $\mathbb{F}_2\h{LOS}$     & $5.4\times10^{-4} \;\rm arcsec^{-1}$ \\
$\mathbb{G}_1\h{LOS}$     & $-4.1\times10^{-3} \;\rm arcsec^{-1}$ & $\mathbb{G}_1\h{LOS}$     & $-1.2\times10^{-3} \;\rm arcsec^{-1}$\\
$\mathbb{G}_2\h{LOS}$     & $-1.6\times10^{-3} \;\rm arcsec^{-1}$ & $\mathbb{G}_2\h{LOS}$     & $-4.8\times10^{-4} \;\rm arcsec^{-1}$ \\
\hline
\end{tabular}
  \caption{LOS parameters which have been used to generate the images in \cref{Fig: rings}. The column \textit{Expected flexion} corresponds to the left panel of \cref{Fig: rings}, which has been generated according to the expected cosmic flexion computed in \cref{Section: flexion value}. The column \textit{Rescaled flexion} corresponds to the right panel of \cref{Fig: rings}, which has been  generated rescaling the flexion used for the left panel by a factor 0.3.}
\label{tab: los params}
\end{table}
We also point out that the difference between the two images in \cref{Fig: rings} is not noticeable by eye, so one cannot necessarily tell \textit{a priori} if the flexion in a given image is small enough to be neglected. However, in the situation when flexion is indeed so tiny that its effect is negligible, as shown in the right panel of \cref{Fig: rings}, the LOS model recovers the correct value of the LOS shear. The posteriors for the LOS shear parameters for this case are shown in the right panel of \cref{Fig: contours LOS}. 

\begin{figure}
    \centering
    \includegraphics[width=0.8\textwidth]{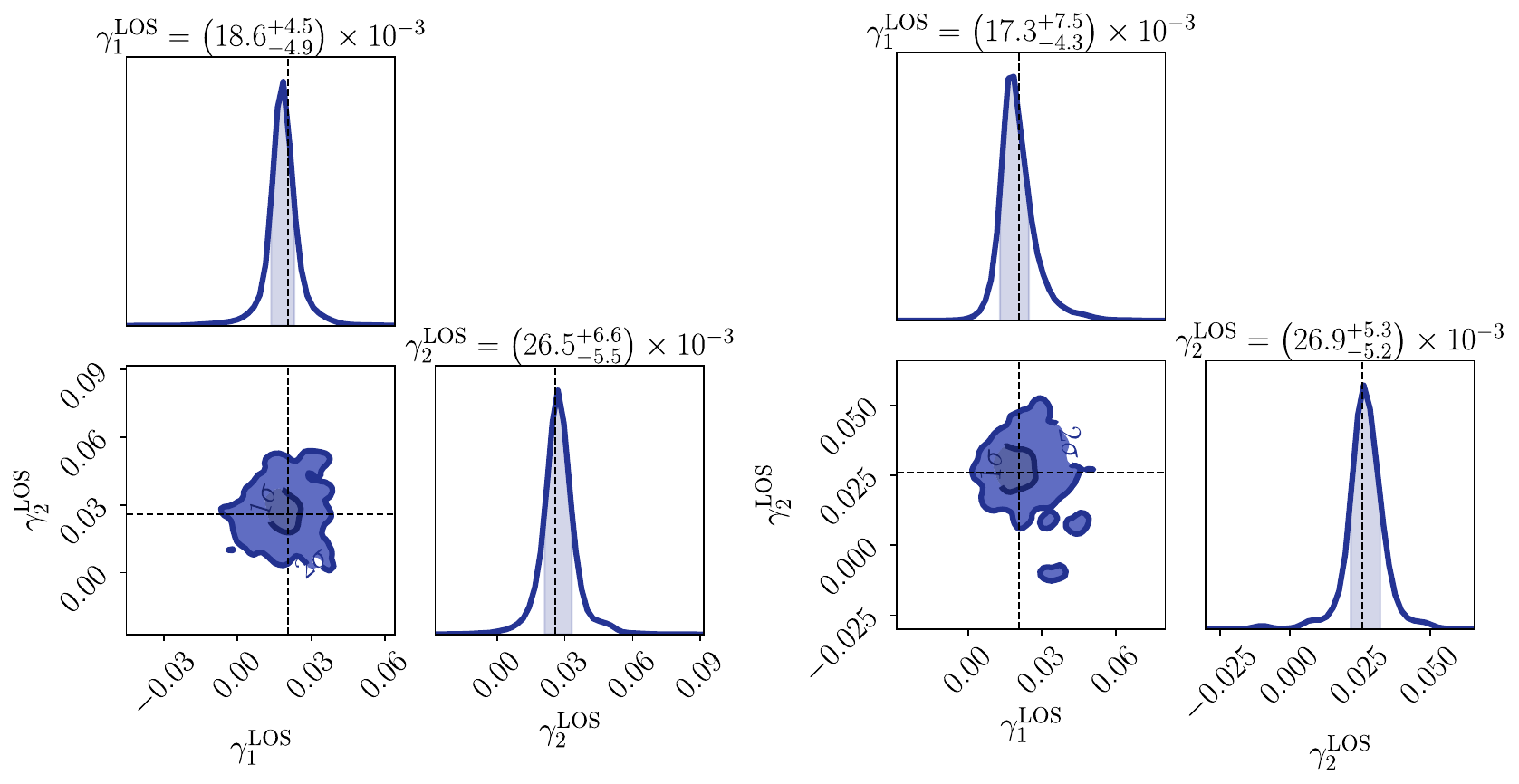}
    \caption{
       \textbf{Inference with the minimal flexion model}, fitting the images in \cref{Fig: rings}, where we give the one- and two-dimensional marginalised posterior distributions of the LOS shear. The dotted lines represent the input value of each parameter, i.e. the one used to generate the image.
    } 
    \label{Fig: contours LOSF}
\end{figure}

We can compare this result with the posteriors obtained using the minimal flexion model, which are shown in \cref{Fig: contours LOSF}. In this case, the LOS shear values are now correctly recovered, even when the flexion is non-negligible. On the other hand, the error bars are larger, due to the larger number of parameters in the minimal flexion model. This means that by taking flexion into account, we can gain accuracy at the cost of losing precision. 

In the case where the flexion is negligible (right panel), the constraints obtained on the LOS shear parameters using the minimal flexion model are less precise than those obtained using the minimal LOS model. This is not surprising; for this type of image, the flexion model has more freedom than necessary, which results in larger error bars. However, in the case where flexion is not negligible, the flexion model is more effective than its shear-only counterpart, as it yields a value of the shear which is in agreement with the fiducial value, while the LOS model does not.

\begin{figure}
    \centering
    \includegraphics[width=\textwidth]{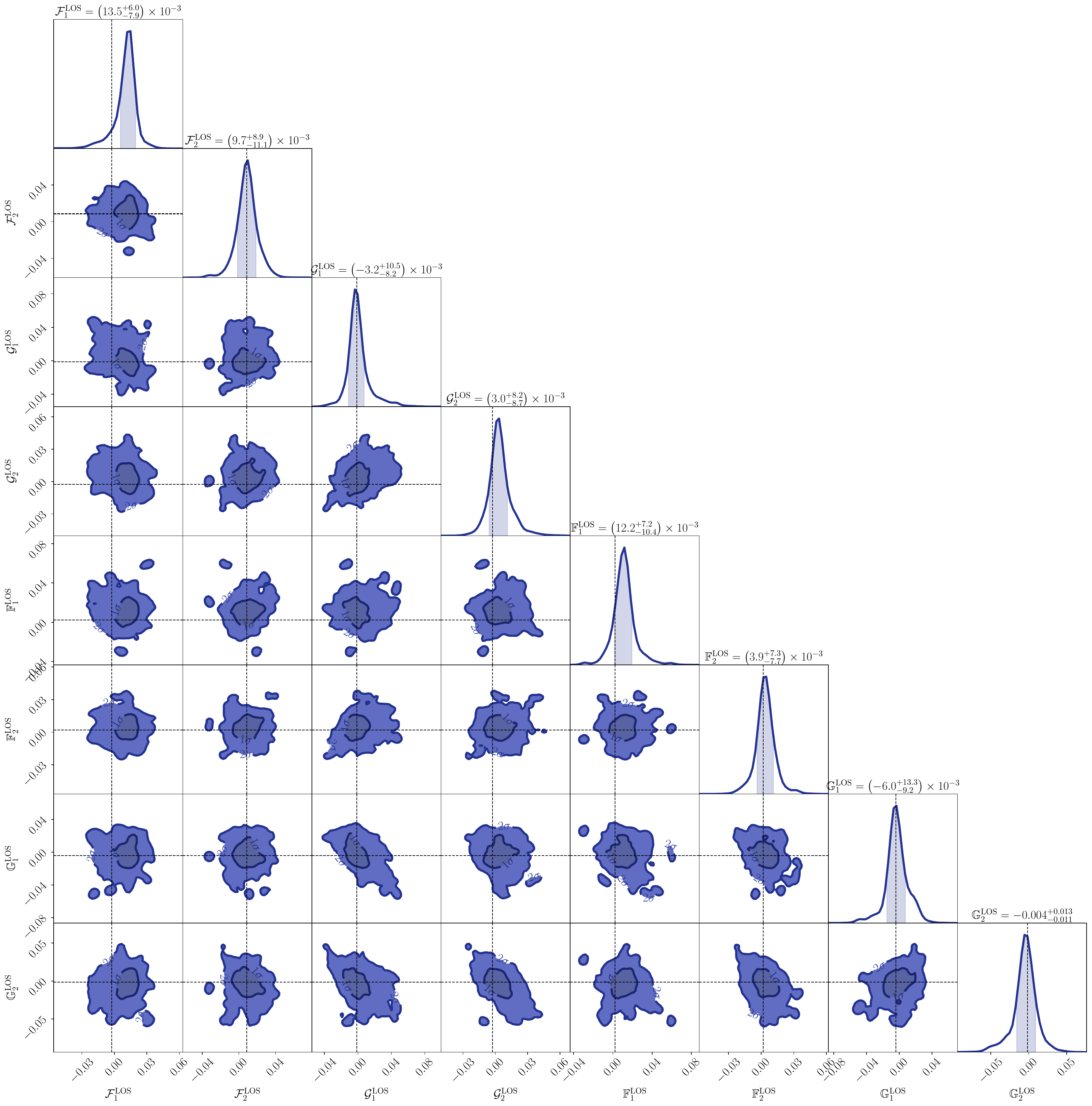}
    \caption{
       \textbf{Inference with the minimal flexion model}, fitting the image in the left panel of \cref{Fig: rings}, where we give the one- and two-dimensional marginalised posterior distributions of the $\F_{\rm LOS}$, $\G_{\rm LOS}$, $\FF$ and $\GG$ flexion parameters. The dotted lines represent the input value of each parameter, i.e. the one used to generate the image. Given the values of the error bars, most parameters are compatibles with 0.
    } 
    \label{Fig: contours LOSF flexion}
\end{figure}

Nevertheless, one should note that in none of these cases are the flexion parameters precisely recovered. The error bars are of the order of $5 \times 10^{-3}-10^{-2} \; \rm arcsec^{-1}$, i.e. larger than the flexion values themselves (see \cref{Fig: contours LOSF flexion} for an illustration). This confirms what we already saw in \cref{Section: min model adv}, and indicates that measuring flexion itself with a single image might be impossible. Our conclusion from this simple test using mock images is that flexion is currently a nuisance parameter which must be taken into account in order to accurately measure the shear, rather than a cosmological observable in its own right. Moreover, because the flexion error bars are of the order of several $10^{-3}\;\rm arcsec^{-1}$, one cannot conclude that flexion is negligible using the minimal flexion model, even though the measured value suggests it is. This means that we cannot try to use the minimal flexion model to pave the way for a more precise measurement using the minimal LOS model.

%% file: Annexe.tex
\appendix

\section{Cosmic flexion computation}
\label{Appendix: cosmic calc}

\subsection{Type-$\G$ flexion}

Here we present the calculation of the cosmic flexion as a function of the matter density contrast $\delta$ along the line of sight. To this end, we can proceed similarly to refs. \cite{Conv_Shear_extended_sources, Disto_Beynd_Shear}, starting by expressing the flexion parameters with point-like perturbers along the line of sight, and then taking the continuous limit. For a point-like lens $l$ with mass $m$ and transverse position $\boldsymbol{y}_l$, we have:
\begin{equation}
    \mathcal{G}_{ilj}(0) = \frac{\partial \gamma_{ilj}}{\partial \underline{\beta}^*_{il}}(0) = - \frac{D_{il}^2 D_{lj}}{D_{ij}} \frac{\partial}{\partial \underline{x}_l^*} \left( \frac{4Gm}{(\underline{x}_l^*-\underline{y}_l^*)^2} \right)\Bigg|_{\underline{x}_l=0} = - \frac{D_{il}^2 D_{lj}}{D_{ij}} \frac{8Gm}{\underline{y}_l^{*3}} \;.
\end{equation}
Using the definition of $\G_{od}$, we get 
\begin{equation}
   \mathcal{G}_{od} (0) = \sum_{l<d}\mathcal{G}_{old}(0) = - \sum_{l<d} \frac{D_{ol}^2 D_{ld}}{D_{od}} \frac{8Gm}{\underline{y}_l^{*3}} \;.
\end{equation}
Now we can take the continuous limit. For a perturber lying at a comoving distance $\chi$ and comoving transverse position $\boldsymbol{\zeta}$, we replace its mass m by ${\rm d}^3m = \overline{\rho}_0 \delta(\eta_0-\chi, \chi, \underline{\zeta})\rm d\chi d^2\boldsymbol{\zeta}$, \footnote{Remember that $\underline\zeta$ is the complex equivalent of $\boldsymbol{\zeta}$, and that a function depending on $\underline\zeta$ also depends on $\underline\zeta^*$. We do not write this dependence for short.} where $\overline{\rho}_0$ denotes the mean matter density today and $\eta_0$ stands for the conformal time today. We thus transform the sum into an integral, to obtain
\begin{equation}\label{GLOS final}
\begin{split}
     \mathcal{G}_{od}(0) &= - 8\pi G\overline{\rho}_0 \int_0^{\chi_s} {\rm d}\chi (1+z)^{-2} W_{od}(\chi) \int_{\mathbb{R}^2} \frac{{\rm d}^2\boldsymbol{\zeta}}{\pi}\delta(\eta_0-\chi, \chi, \underline{\zeta}) \left(\frac{1+z}{\underline{\zeta}^*}\right)^3 \\
     &= - 3 \Omega_{\rm m,0} H_0^2 \int_0^{\chi_s} {\rm d}\chi (1+z) W_{od}(\chi) \int_{\mathbb{R}^2} \frac{\rm d^2\boldsymbol{\zeta}}{\pi \zeta^3} \rm e^{3i\phi}\delta(\eta_0-\chi, \chi, \underline{\zeta}),
\end{split}
\end{equation}
where $\underline{\zeta} = \zeta \rm e^{i\phi}$ and the weight function $W_{od}$ is given by
\begin{equation}
    W_{od}(\chi) = \left\{ 
    \begin{array}{ll}
        \frac{f_K(\chi_d - \chi) f_K(\chi)^2}{f_K(\chi_d)} & \mbox{if} \; 0 \leq \chi \leq \chi_d \;, \\
        0 & \mbox{otherwise.}
    \end{array}
    \right.
\end{equation}
Using the same procedure allows one to write the various type-$\G$ flexion parameters. They always have the same expression except that the weight function changes according to
\begin{align}
    W_{os}(\chi) &= \left\{ 
    \begin{array}{ll}
        \frac{f_K(\chi_s - \chi) f_K(\chi)^2}{f_K(\chi_s)} & \mbox{if} \; 0 \leq \chi \leq \chi_s \;, \\
        0 & \mbox{otherwise,}
    \end{array}
    \right. \\
    W_{1ds}(\chi) &= \left\{ 
    \begin{array}{ll}
        \frac{ f_K(\chi - \chi_d)^2 f_K(\chi_s - \chi)f_K(\chi_s)}{f_K(\chi_s - \chi_d)^2} & \mbox{if} \; \chi_d \leq \chi \leq \chi_s \;, \\
        0 & \mbox{otherwise,}
    \end{array}
    \right. \\
    W_{2ds}(\chi) &= \left\{ 
    \begin{array}{ll}
        \frac{f_K(\chi) f_K(\chi - \chi_d) f_K(\chi_s - \chi)}{f_K(\chi_s - \chi_d) }  & \mbox{if} \; \chi_d \leq \chi \leq \chi_s \;, \\
        0 & \mbox{otherwise,}
    \end{array}
    \right. \\
    W_{\rm LOS} (\chi) &= \left\{ 
    \begin{array}{ll}
        f_K(\chi)^2 \left[\frac{f_K(\chi_d - \chi)}{f_K(\chi_d)} + \frac{f_K(\chi_s - \chi)}{f_K(\chi_s)}\right] & \mbox{if} \; 0 \leq \chi \leq \chi_d \;, \\
        f_K(\chi)f_K(\chi_s - \chi) \left[ \frac{f_K(\chi)}{f_K(\chi_s)} - \frac{f_K(\chi - \chi_d)}{f_K(\chi_s - \chi_d)} \right] & \mbox{if} \; \chi_d \leq \chi \leq \chi_s \;, \\
        0 & \mbox{otherwise,}
    \end{array}
    \right. \\
    W_{\mathbb{LOS}}(\chi) &= \left\{ 
    \begin{array}{ll}
        \frac{f_K(\chi_d - \chi) f_K(\chi)^2}{f_K(\chi_d)} & \mbox{if} \; 0 \leq \chi \leq \chi_d \;, \\
        \frac{ f_K(\chi - \chi_d) f_K(\chi_s - \chi)}{f_K(\chi_s - \chi_d) } \left[ \frac{f_K(\chi - \chi_d) f_K(\chi_s)}{f_K(\chi_s - \chi_d)} - f_K(\chi) \right] & \mbox{if} \; \chi_d \leq \chi \leq \chi_s \;, \\
        0 & \mbox{otherwise.}
    \end{array}
    \right.
\end{align}

\begin{figure}[h]
    \centering
    \includegraphics[width=\textwidth]{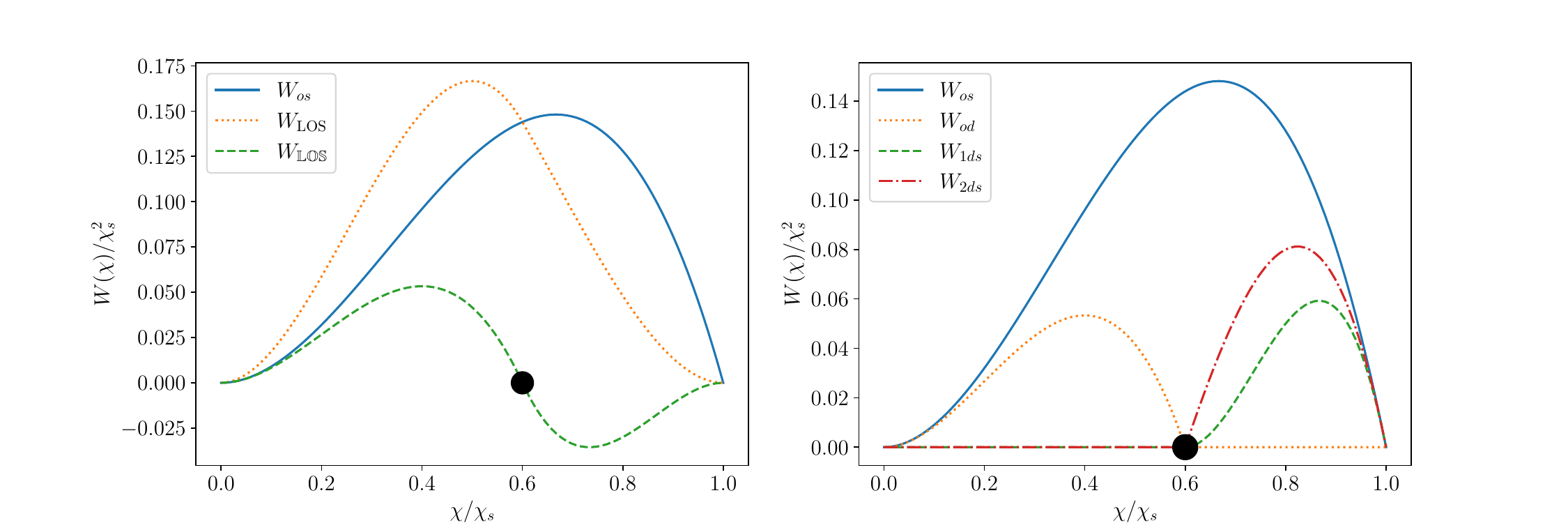}
    \caption{Weight functions of the different types of flexion. The standard cosmic flexion weight function corresponds to $W_{os}$. Here we have taken $K=0$ so that $f_K(\chi) = \chi$ and the main lens is at $\chi_d=0.6\chi_s$.}
    \label{Fig: Weight functions}
\end{figure}

\noindent
Note that $W_{os}$ corresponds to the standard weak lensing weight function. We note that we indeed recover the results from \cite{Bacon_2006} and that both LOS weight functions differ from the standard weak lensing one. An illustration is given in \cref{Fig: Weight functions}.
We see that the minus sign which comes with $^{(2)}\mathcal{G}_{ds}$ in the definition of $\GG$ -- associated with the fact that $^{(2)}\mathcal{G}_{ds}$ has a superior contribution to $^{(1)}\mathcal{G}_{ds}$ -- leads the background to have a contribution with opposite sign to the foreground. For $\G_{\rm LOS}$ this is quite different. Compared to the standard weak lensing situation, the presence of $-^{(2)}\G_{ds}$ decreases the weight function in the background, but $\G_{od}$ adds up in the foreground, resulting in a high valued weight function whose maximum is displaced towards the foreground.

\subsection{Type-$\F$ flexion}

The calculation of type-$\F$ flexion is a little more tricky. Here, we cannot use the point-like lens as an intermediate step as the type-$\mathcal{F}$ flexion of such a lens is zero.\footnote{Indeed, $\mathcal{F}_{ilj} = -D_{il}^2/(D_{lj}D_{ij}) \partial/\partial \underline{x}_l \left( 4Gm(\underline{x}_l^*-\underline{y}_l^*)^{-2} \right) = 0 $.}
However, the cosmic type-$\mathcal{F}$ flexion is non-zero due to the collective effect of all the masses along the line of sight. In this regard, the case of type-$\mathcal{F}$ flexion is very similar to the convergence. Also as in the case of the convergence, it is the mass inside a light beam that is responsible for the flexion,  not the mass outside of the beam (see \cite{Conv_Shear_extended_sources} for more details). This leads us to try another approach, closer to the one used to derive the cosmic convergence. Here we will use the fact that $\mathcal{F}_{ilj}=\partial \kappa_{ilj}/\partial \underline{\beta}_{il}^*$. Let us start by calculating $\mathcal{F}_{od}$:
\begin{equation}
    \mathcal{F}_{od} = \sum_{l<d} \mathcal{F}_{old} = \sum_{l<d} \frac{\partial \kappa_{old}}{\partial \underline{\beta}^*_{ol}} \approx \frac{\partial }{\partial \underline{\theta}^*} \sum_{l<d} \kappa_{old} \;.
\end{equation}
Taking the continuous limit, the sum corresponds to the cosmic convergence along the line of sight between the observer and a source located at $\chi_d$. Using equation (6.16) of \cite{Bartelmann_2001}, we find:
\begin{equation} \label{cosmic F}
\begin{split}
    \mathcal{F}_{od}(0) &= \frac{\partial }{\partial \underline{\theta}^*} \left( \frac{3}{2}\Omega_{\rm m,0}H_0^2 \int_0^{\chi_d} (1+z)\frac{f_K(\chi_d - \chi) f_K(\chi)}{f_K(\chi_d)} \delta[\eta_0-\chi, \chi, f_K(\chi)\underline{\theta}] \rm d\chi  \right) \Bigg|_{\underline{\theta} = 0}\\
    &= \frac{3}{2}\Omega_{\rm m,0} H_0^2 \int_0^{\chi_s} (1+z) W_{od}(\chi) \Delta(\eta_0-\chi, \chi, 0) \rm d\chi \;,
\end{split}
\end{equation}
where $\Delta \define \partial \delta/\partial \underline\zeta^*$ is the complex equivalent of half of the gradient of the matter density contrast across a plane orthogonal to the line of sight. For the other type-$\F$ flexion parameters, only the prefactor in front of the convergence changes, which does not affect the way the convergence is computed. Therefore, equation (6.16) of \cite{Bartelmann_2001} is still valid, up to some geometric factors. As for type-$\G$ flexion, we find that all type-$\F$ flexion parameters have the same expressions (given by \cref{cosmic F}) up to the weight function, which are the same as the weight functions for $\G$ given before.

The method used here to compute type-$\F$ flexion, based on the derivation of $\kappa$ \textit{after} it is computed in the continuous limit, can be used to write type-$\G$ in another form. Following the same steps, we get
\begin{equation}
    \mathcal{G}(0) = - \frac{3}{2} \Omega_{\rm m,0} H_0^2 \int_0^{\chi_s} {\rm d}\chi (1+z) W(\chi) \int_{\mathbb{R}^2} \frac{\rm d^2\boldsymbol{\zeta}}{\pi \zeta^2} \rm e^{2i\phi}\Delta(\eta_0-\chi, \chi, \underline{\zeta}),
\end{equation}
where $\G$ can be any of the type-$\G$ flexion parameters and $W$ is the corresponding weight function. We can then compare this expression with the one obtained with the first method \cref{GLOS final}, which gives us the identity
\begin{equation}
    \frac{1}{2}\int_{\mathbb{R}^2} \frac{\rm d^2\boldsymbol{\zeta}}{\pi \zeta^2} \rm e^{2i\phi}\Delta(\eta_0-\chi, \chi, \underline{\zeta}) = \int_{\mathbb{R}^2} \frac{\rm d^2\boldsymbol{\zeta}}{\pi \zeta^3} \rm e^{3i\phi}\delta(\eta_0-\chi, \chi, \underline{\zeta}) \;.
\end{equation}
This can easily be checked using integration by parts in Cartesian coordinates. We thus have two equivalent ways to expression the type-$\mathcal{G}$ flexion, one expressed in terms of the density contrast $\delta$ and the other in terms of its derivative $\Delta$.

\section{Cosmic flexion variance calculation}

\subsection{First definition}
\label{Appendix: variance1}

Here we present the calculation of the flexion variance. Starting with \cref{F eff} and using the expression of $\F$ given in \cref{Final cosmic F}:
\begin{equation}
    \mathcal{F}_{\rm eff}(\underline\vartheta) = \frac{3}{2}\Omega_{\rm m,0} H_0^2 \int {\rm d}^2\boldsymbol\Pi \; p(\boldsymbol{\Pi}) \int_{0}^{\chi_s} {\rm d}\chi (1+z) W(\chi, \boldsymbol\Pi) \Delta[\eta_0-\chi, \chi, f_K(\chi)\underline\vartheta] \;.
\end{equation}
To compute the correlation function, we will introduce the Fourier transform of the matter density contrast. Let us label $\boldsymbol x$ the spatial position corresponding to the comoving position $\chi$ and transverse position $\underline\zeta = f_K(\chi)\underline\vartheta$, and $\eta = \eta_0 - \chi$ the conformal time at the corresponding comoving distance. Then:
\begin{equation}
    \Delta(\eta, \boldsymbol x) = \frac{\partial \delta}{\partial \underline\zeta^*} = \frac{1}{2}\left( \frac{\partial \delta}{\partial \zeta_1} + \rm i \frac{\partial \delta}{\partial \zeta_2} \right) = \int \frac{\rm d^3\boldsymbol k}{(2\pi)^3} \hat{\delta}(\eta, \boldsymbol k) {\rm i}\frac{k_1+{\rm i}k_2}{2} {\rm e}^{{\rm i}\boldsymbol k \boldsymbol x} \;.
\end{equation}
In the same way, we need to express the conjugate of $\Delta$, which is a complex number. However, keeping in mind that the density contrast $\delta$ is real, we get
\begin{equation}
    \Delta(\eta, \boldsymbol x)^* = \frac{1}{2}\left( \frac{\partial \delta^*}{\partial \zeta_1} - \rm i \frac{\partial \delta^*}{\partial \zeta_2} \right) = \frac{1}{2}\left( \frac{\partial \delta}{\partial \zeta_1} - \rm i \frac{\partial \delta}{\partial \zeta_2} \right) = \int \frac{{\rm d}^3\boldsymbol l}{(2\pi)^3} \hat{\delta}(\eta, \boldsymbol l) {\rm i}\frac{l_1- {\rm i} l_2}{2} {\rm e}^{{\rm i}\boldsymbol l \boldsymbol x} \;.
\end{equation}
Then, using $<\hat{\delta}(\eta_1, \boldsymbol k)\hat{\delta}(\eta_2, \boldsymbol l)> = (2\pi)^3\delta_{\rm D}(\boldsymbol k + \boldsymbol l) P_{\delta}(\eta_1, \eta_2, k)$, where $P_{\delta}$ is the power spectrum of the density contrast, we get
\begin{equation}
    \langle\Delta(\eta_1, \boldsymbol x_1)\Delta(\eta_2, \boldsymbol x_2)^*\rangle = \int \frac{{\rm d}^3\boldsymbol k}{(2\pi)^3} \frac{k_1^2+k_2^2}{4} {\rm e}^{{\rm i}\boldsymbol k (\boldsymbol x_1 - \boldsymbol x_2)} P_{\delta}(\eta_1, \eta_2, k) \;.
\end{equation}
Using Limber's approximation (\cite{Limber1953}) we can further simplify this expression:
\begin{equation} \label{DeltaDelta* after Limber}
    \langle \Delta(\eta_1, \boldsymbol x_1)\Delta(\eta_2, \boldsymbol x_2)^* \rangle \approx \frac{\delta_{\rm D}(\chi_1 - \chi_2)}{4 f_K(\chi_1)^4}\int \frac{\rm d^2\boldsymbol \ell}{(2\pi)^2} \ell^2 {\rm e}^{{\rm i}\boldsymbol\ell (\boldsymbol\vartheta_1-\boldsymbol\vartheta_2)} P_{\delta}\left(\eta_1, \frac{\ell}{f_K(\chi_1)} \right) \;.
\end{equation}
As we aim to compute the variance, $\boldsymbol\vartheta_1 = \boldsymbol{\vartheta}_2$, and therefore this allows us to write
\begin{equation}
    \sigma^2_{\mathcal{F}} = \frac{1}{4}\left(\frac{3}{2}\Omega_{\rm m,0} H_0^2\right)^2  \int_{0}^{\chi_s} {\rm d}\chi (1+z)^2 q_{\rm F}(\chi)^2 \int \frac{\rm d^2\boldsymbol \ell}{(2\pi)^2} \ell^2 P_{\delta}\left(\eta, \frac{\ell}{f_K(\chi)} \right),
\end{equation}
with
\begin{equation}
    q_{\rm F}(\chi) = \int {\rm d}^2\boldsymbol\Pi \; p(\boldsymbol{\Pi}) \frac{W(\chi, \boldsymbol\Pi)}{f_K(\chi)^2} \;.
\end{equation}

\noindent
We can now decompose the integral over $\boldsymbol\ell$ in polar coordinates. The final result reads
\begin{equation} \label{sigma1-app}
    \sigma_\F^2 = \sigma_\G^2 = \int_0^{\infty} \frac{\ell^3}{2\pi} P_{\rm F}(\ell) \rm d\ell \;,
\end{equation}
with
\begin{equation}
    P_{\rm F}(\ell) = \frac{1}{4}\left(\frac{3}{2}\Omega_{\rm m,0} H_0^2\right)^2  \int_{0}^{\chi_s} {\rm d}\chi (1+z)^2 q_{\rm F}(\chi)^2 P_{\delta}\left(\eta_0 - \chi, \frac{\ell}{f_K(\chi)} \right) \;.
\end{equation}

\subsection{Averaged flexion}
\label{Appendix: variance2}

In this subsection we compute the cosmic flexion variance but with the alternative definition
\begin{equation}
    \mathcal{F}_{\rm eff}(\underline{\vartheta}) = \frac{1}{\pi\theta_S^2} \int_S {\rm d}^2\boldsymbol{s} \int {\rm d}^2\boldsymbol\Pi \; p(\boldsymbol{\Pi}) \mathcal{F}(\underline{s}, \chi_d, \chi_s) \;,
\end{equation}

\noindent
with $S$ being the surface defined by the disk on the flat sky centred in $\underline\vartheta$ and of radius $\theta_S$. The beginning of the calculation is the same as in \cref{Appendix: variance1} up to \cref{DeltaDelta* after Limber}, thus we get:
\begin{equation} \label{sigma smoothed inter}
    \sigma^2_{\mathcal{F}} = \frac{1}{4}\left(\frac{3}{2}\Omega_{\rm m,0} H_0^2\right)^2  \int_{0}^{\chi_s} {\rm d}\chi (1+z)^2 q_{\rm F}(\chi)^2 \int \frac{\rm d^2\boldsymbol \ell}{(2\pi)^2} \ell^2 P_{\delta}\left(\eta, \frac{\ell}{f_K(\chi)} \right) I(\boldsymbol\ell),
\end{equation}
with
\begin{equation}
    I(\boldsymbol\ell) = \frac{1}{\pi^2\theta_S^4}\int_{S_1\times S_2} {\rm e}^{{\rm i}\boldsymbol{\ell}(\boldsymbol{s}_1 - \boldsymbol{s}_2)} {\rm d}^2\boldsymbol{s}_1 {\rm d}^2\boldsymbol{s}_2 \;.
\end{equation}

\noindent As we are computing the variance, $\boldsymbol\vartheta_1 = \boldsymbol\vartheta_2$, and thus $S_1 = S_2$ (but of course $\boldsymbol{s}_1 \neq \boldsymbol{s}_2$). In $I$, $\boldsymbol{s}_1$ and $\boldsymbol{s}_2$ are decoupled, so we can split the integral into the product of two integrals, and with a change of variable one can show one is equal to the other, yielding
\begin{equation}
    I(\boldsymbol{\ell}) = \left( \frac{1}{\pi\theta_S^2} \int_S {\rm e}^{{\rm i}\boldsymbol{\ell}\boldsymbol{s}} {\rm d}^2\boldsymbol{s} \right)^2 \;.
\end{equation}

\noindent We then decompose the integral in polar coordinates $(s, \varphi)$. We can always define those coordinates so that $\boldsymbol{\ell} \boldsymbol{s} = \ell s \cos(\varphi)$. With a change of variable we then recognise a Bessel function, which gives
\begin{equation}
\begin{split}
    I(\boldsymbol{\ell}) &= \left( \frac{2\pi}{\pi\theta_S^2} \int_0^{\theta_S} s J_0(\ell s) {\rm d}s \right)^2 \\
    &= \left( \frac{2}{\ell^2\theta_S^2} \int_0^{\ell\theta_S} x J_0(x) {\rm d}x \right)^2 \\
    &= \left( \frac{2}{\ell^2\theta_S^2} \Big[x J_1(x)\Big]_{x=0}^{x=\ell\theta_S}  \right)^2 \\
    &= \left[ \frac{2}{\ell\theta_S} J_1(\ell\theta_S) \right]^2 \;.
\end{split}
\end{equation}

\noindent Notice that $I(\boldsymbol{\ell})$ in fact only depends on the modulus of $\boldsymbol{\ell}$. Inserting this into \cref{sigma smoothed inter} and using the definitions from \cref{Appendix: variance1}, we finally get
\begin{equation}
    \sigma_\F^2 = \sigma_\G^2 = \int_0^{\infty} \frac{\ell^3}{2\pi} P_{\rm F}(\ell) \left[ \frac{2 J_1(\ell\theta_S)}{\ell\theta_S} \right]^2 \rm d\ell \;.
\end{equation}

\noindent Note that unlike the integral in \cref{sigma1-app}, this one converges. Indeed, for high $\ell$ we extrapolate $P_{\rm F}(\ell) \in \Theta(l^{-3})$ and $J_1(\ell\theta_S)/\ell \in \Theta(l^{-3/2})$ for $\ell \to \infty$, so the integrand is of order $\ell^{-3}$ which means the integral converges.